\documentclass[10pt]{article}
\usepackage{graphicx}
\usepackage{amsmath}
\usepackage{amssymb}
\usepackage{subfigure}
\usepackage{url}
\usepackage[round]{natbib}

\def\reals{\mathbb{R}}
\def\w{\mathbf{w}}
\def\R{\pmb{R}}
\def\1{\mathbf{1}}
\def\0{\mathbf{0}}
\def\r{\mathbf{r}}
\def\mmu{\pmb{\mu}}
\def\trp{^{\top}}
\def\w{\mathbf{w}}
\def\y{\mathbf{y}}
\def\bb{\mathbf{b}}
\def\uu{\mathbf{u}}
\def\aa{\mathbf{a}}

\newcommand{\vv}{\mathbf{v}}
\newcommand{\sss}{\mathbf{s}}
\newcommand{\llambda}{\pmb{\lambda}}
\newcommand{\A}{\pmb{A}}

\begin{document}

\title{Sparse and stable Markowitz portfolios}

\author{Joshua Brodie, Princeton University, Applied Mathematics\\
Ingrid Daubechies, Princeton University, Math. Dept. and PACM\\
Christine De Mol,
Universit\'e Libre de Bruxelles, Math. Dept. and ECARES \\
Domenico Giannone, European Central Bank, ECARES and CEPR\\
Ignace Loris, Mathematics Dept., Vrije Universiteit Brussel }
\date{May 28, 2008}

\maketitle

\begin{abstract}
We consider the problem of portfolio selection within the classical
Markowitz mean-variance framework, reformulated as a constrained
least-squares regression problem.  We propose to add to the
objective function a penalty proportional to the sum of the absolute
values of the portfolio weights. This penalty regularizes
(stabilizes) the optimization problem, encourages sparse portfolios
(i.e. portfolios with only few active positions), and allows to
account for transaction costs. Our approach recovers as special
cases the no-short-positions portfolios, but does allow for short
positions in limited number. We implement this methodology on two
benchmark data sets constructed by Fama and French. Using only a
modest amount of training data, we construct portfolios whose
out-of-sample performance, as measured by Sharpe ratio, is
consistently and significantly better than that of the na\"ive
evenly-weighted portfolio which constitutes, as shown in recent
literature, a very tough benchmark.
\end{abstract}

\begin{description}
\item[Keywords: ] Portfolio Choice, Sparsity, Penalized Regression
\end{description}

\section{Introduction}

In 1951, Harry Markowitz ushered in the modern era of portfolio
theory by applying simple mathematical ideas to the problem of
formulating optimal investment portfolios \citep{Mar52}. Single
minded pursuit of high returns constitutes a poor strategy,
Markowitz argued.  Instead, he suggested, rational investors must
balance a desire for high returns with a desire for low risk, as
measured by variability of returns.

It is not trivial, however, to translate Markowitz's conceptual
framework into a satisfactory portfolio selection algorithm in a
real-world context. Indeed, in a recent survey, \citet{DeM07}
examined several portfolio construction algorithms inspired by the
Markowitz framework.  The authors found that, given a reasonable
amount of training data, none of the surveyed algorithms is able to
significantly or consistently outperform the na\"{i}ve strategy in
which each available asset is given an equal weight in the
portfolio. This disappointing algorithm performance is likely due,
at least in part, to the structure of the Markowitz optimization
framework as originally proposed.  Specifically, the optimization at
the core of the Markowitz scheme is, as originally formulated,
empirically unstable: small changes in assumed asset returns,
volatilities or correlations can have large effects on the output of
the optimization procedure.  In this sense, the classic Markowitz
portfolio optimization can be viewed as an ill-posed (or
ill-conditioned) inverse problem. Such problems are frequently
encountered in many other fields, where a variety of regularization
procedures have been proposed to tame the troublesome instabilities
\citep{BerBoc98}.

In this paper, we discuss a regularization of Markowitz's portfolio construction.
We shall restrict ourselves to the traditional Markowitz mean-variance approach\footnote{
Similar ideas
could also be applied to different portfolio construction frameworks
considered in the literature.};
moreover, we focus on one particular regularization method,
and highlight some very special
properties of the regularized portfolios obtained through its use.

Our proposal consists of adding an $\ell_1$ penalty to the Markowitz
objective function.  We allow ourselves to adjust the importance of
this penalty  with a ``tunable'' coefficient. For large values of
this coefficient, optimization of the penalized objective function
turns out to be equivalent to solving the original (unpenalized)
problem under an additional positivity condition on the weights. As
the tunable coefficient is decreased, the optimal solutions are
given more and more latitude to include short positions. The optimal
solutions for our penalized objective function can thus be seen as
natural generalizations of the ``no-short-positions'' portfolios
considered by  \citet{JaMa03}. In addition to stabilizing the
optimization problem \citep{Dau04} and generalizing
no-short-positions-constrained optimization, the $\ell_1$ penalty
facilitates treatment of transaction costs.  For large investors,
whose principal cost is a fixed bid-ask spread, transaction costs
are effectively proportional to the gross market value of the
selected portfolio, and thus to our $\ell_1$ penalty term. For small
investors, volume independent ``overhead'' costs cannot be ignored,
and thus transaction costs are best modeled via a combination of an
$\ell_1$ penalty term and the number of assets transacted;
minimizing such a combination is tantamount to searching for sparse
solutions (sparse portfolios or sparse changes to portfolios), a
goal that we shall see is also achieved by our use of an $\ell_1$
penalty term.\footnote{A {\it sparse} portfolio is a portfolio with
few active positions, i.e. few non-zero weights.}

We implement the methodology and compute efficient investment
portfolios using as our assets two sets of portfolios constructed by
Fama and French: the 48 industry portfolios and the 100 portfolios
formed on size and book-to-market. Using data from 1973 to 2006, we
construct an ensemble of portfolios for various values of our
tunable coefficient and track their out-of-sample performances. We
find a consistent and significant increase in Sharpe ratio compared
to the na\"{i}ve equal-weighting strategy. When using the 48
industry portfolios as our assets, we find that the best portfolios
we construct have no short positions. When our assets are the 100
portfolios formed on size and book-to-market, we find that the best
portfolios constructed by our methodology {\em do} include short
positions.

We are not alone in proposing the use of regularization in the
context of Markowitz-inspired portfolio construction; \cite{DGNU}
discuss several different regularization techniques for the
portfolio construction problem, including the imposition of
constraints on appropriate norms of the portfolio weight vector. Our
work\footnote{The first presentation of our work is given in
\cite{BDDG}, independent of and simultaneous with \cite{DGNU}.}
differs from \cite{DGNU} in that our goal is not only
regularization: we are interested in particular in the stable
construction of sparse portfolios, which is achieved by $\ell_1$
penalization, as demonstrated by our analysis and examples.

The organization of our paper is as follows. In the next Section, we
formulate the problem of portfolio selection and we describe our
methodology based on $\ell_1$-penalized least-squares regression. In
Section 3, we describe key mathematical properties of the portfolios
we construct and devise an efficient algorithm for computing them.
In Section 4, we present empirical results consisting of an
out-of-sample performance evaluation of our sparse portfolios. In
Section 5, we discuss possible extensions of our methodology to
other portfolio construction problems which can be naturally
translated into optimizations involving $\ell_1$ penalties. Finally,
Section 6 summarizes our findings.

\section{Sparse portfolio construction}

We consider $N$ securities and denote their returns at time $t$ by
$r_{i,t}$, $i=1, \dots, N$. We write $\mathbf{r}_t = (r_{1,t},
\ldots, r_{N,t})\trp$ for the $N\times 1$ vector of returns at time
$t$. We assume that the returns are stationary and write
$\mathbf{E}[\r_t] = \mmu $ for the vector of expected returns of the
different assets, and $\mathbf{E}[(\r_t- \mmu)(\r_t-\mmu)\trp]=
\pmb{C}$ for the covariance matrix of returns.

A portfolio is defined to be a list of weights $w_i$ for assets $i
= 1, \ldots, N$ that represent the amount of capital to be
invested in each asset. We assume that one unit of capital is
available and require that capital to be fully invested.  Thus we
must respect the constraint that $\sum_{i=1}^N w_i = 1$. We
collect the weights in an $N \times 1$ vector $\w=(w_1, \ldots,
w_N)\trp$. The normalization constraint on the weights can thus be
rewritten as  $ \w\trp \1_N= 1$, where $\1_N$ denotes the $N
\times 1$ vector in which every entry is equal to 1.  For a given
portfolio $\w$, the expected return and variance are equal to
$\w\trp\mmu$ and $\w\trp\pmb{C}\w$, respectively.

In the traditional Markowitz portfolio optimization, the objective
is to find a portfolio which has minimal variance for a given
expected return $\rho= \w\trp \mmu$. More precisely, one seeks
$\widetilde{\w}$ such that:
\begin{eqnarray*}
\widetilde{\w} & = & \arg \min_{\w} \w\trp\pmb{C}\w \\
& \mbox{s. t.} &  \w\trp\mmu = \rho\\
& & \w\trp \1_N= 1.
\end{eqnarray*}

Since $\pmb{C} = \mathbf{E}[\r_t\r\trp_t]\, - \, \mmu\mmu\trp$, the
minimization problem is equivalent to
\begin{eqnarray*}
\widetilde{\w} & = & \arg \min_{\w}
\mathbf{E}\left[|\rho-\w\trp\r_t|^2\right] \\
& \mbox{s. t.} &  \w\trp\mmu = \rho\\
& & \w\trp \1_N= 1.
\end{eqnarray*}

For the empirical implementation, we replace expectations with
sample averages. We set $\widehat{\mmu}=\frac{1}{T}\sum_{t=1}^T
\r_t$ and define  $\pmb{R}$ to be the $T \times N$ matrix of which
row $t$ is given by $\r_t\trp$, that is, $\pmb{R}_{t,i} =
\left(\r_t\right)_i = r_{i,t}$.  Given this notation, we thus seek
to solve the following optimization problem
\begin{eqnarray}
\widehat{\w} & = &
\arg \min_{\w} \frac{1}{T} \|\rho\1_T-\pmb{R}\w\|^2_2\label{regression}\\
& \mbox{s. t.} &  \w\trp\widehat{\mmu}= \rho \nonumber\\
& & \w\trp \1_N= 1,\nonumber
\end{eqnarray}
where, for a vector $\aa$ in $\reals^T$, we denote by
$\|\aa\|^2_2$ the sum $\sum_{t=1}^T \aa_t^2$.

This problem requires the solution of a constrained multivariate
regression involving many potentially collinear variables.  While
this problem is analytically simple, it can be quite challenging in
practice, depending on the nature of the matrix $\pmb{R}$.
Specifically, the condition number --- defined to be the ratio of
the largest to smallest singular values of a matrix --- of $\pmb{R}$
can effectively summarize the difficulty we will face when trying to
perform this optimization in a stable way. When the condition number
of $\pmb{R}$ is small, the problem is numerically stable and easy to
solve.  However, when the condition number is large, a
non-regularized numerical procedure will amplify the effects of
noise anisotropically, leading to an unstable and unreliable
estimate of the vector $\mathbf{w}$.  As asset returns tend to be
highly correlated, the smallest singular value of $\pmb{R}$ can be
quite small, leading to a very large condition number and thus very
unstable optimizations in a financial context. It is this sort of
instability that likely plagues many of the algorithms reviewed by
\citet{DeM07} and renders them unable to outperform the na\"{i}ve
portfolio.

To obtain meaningful, stable results for such ill-conditioned
problems, one typically adopts a {\em regularization} procedure. One
fairly standard approach is to augment the objective function of
interest with a penalty term, which can take many forms and ideally
should have a meaningful interpretation in terms of the specific
problem at hand. We propose here to add a so-called $\ell_1$ penalty
to the original Markowitz objective function (\ref{regression}). We
thus seek to find a vector of portfolio weights $\w$ that minimizes
\begin{eqnarray}
\w^{[\tau]} & = &
\arg \min_{\w}\left[ ||\rho \1_T - \pmb{R} \w||^2_2+ \tau||\w||_1\right]\label{lasso}\\
& \mbox{s. t.} &  \w\trp\widehat{\mmu}= \rho \label{total_return_constraint}\\
& & \w\trp \1_N= 1.\label{total_weight_constraint}
\end{eqnarray}

Here the $\ell_1$ norm of a vector $\w$ in $\reals^N$ is defined by
$\|\w\|_1:=\sum_{i=1}^N |w_i|~$, and $\tau$ is a parameter that
allows us to adjust the relative importance of the $\ell_1$
penalization in our optimization. Note that we absorbed the factor
$1/T$  from (\ref{regression}) in the parameter $\tau$. The
particular problem of minimizing an (unconstrained) objective
function of the type (\ref{lasso}) was named {\em lasso regression}
by \citet{Tib96}.

Adding an $\ell_1$ penalty to the objective function
(\ref{regression}) has several useful consequences:
\begin{itemize}
\item{It promotes {\em sparsity}. That penalizing or minimizing
$\ell_1$ norms can have a sparsifying effect has long been
observed in statistics (see e.g. \citet{Che01} and the references
therein). Minimization of $\ell_1$ penalized objective functions
is now a widely used technique when sparse solutions are
desirable.  Sparsity should also play a key role in the task of
formulating investment portfolios. Indeed, investors frequently
want to be able to limit the number of positions they must create,
monitor and liquidate.  By considering suitably large values of
$\tau$ in (\ref{lasso}), one can achieve just such an effect
within our framework. Figure \ref{fig:sparsity} illustrates
geometrically how the addition of an $\ell_1$ term to the
unconstrained volatility minimization encourages sparse
solutions.}

\begin{figure}
\centering \subfigure[Very small $\ell_1$ penalty: tangency on
edge.]{\includegraphics[scale = .2]{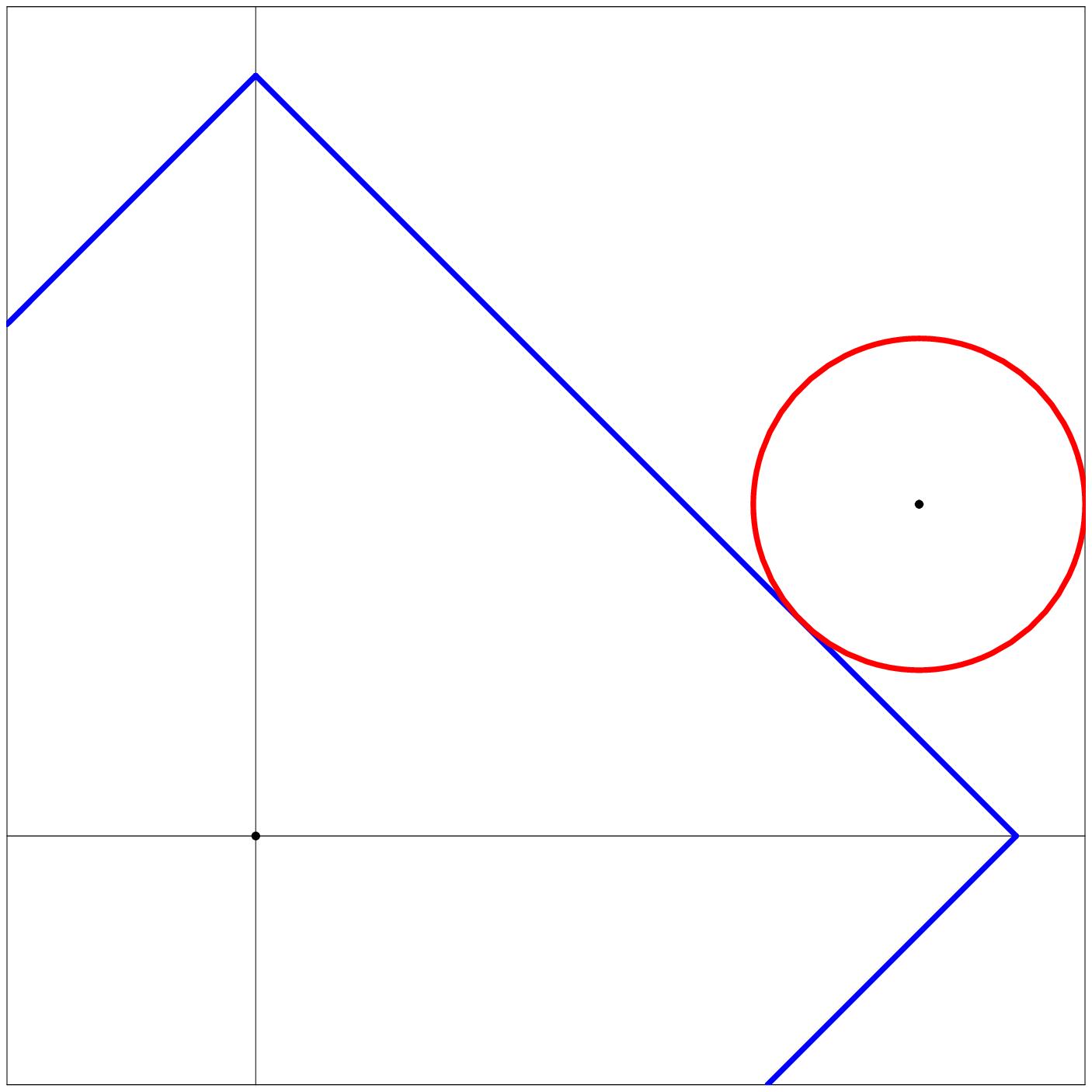}}\quad
\subfigure[Small $\ell_1$ penalty: tangency on edge, nearing
vertex.]{\includegraphics[scale = .2]{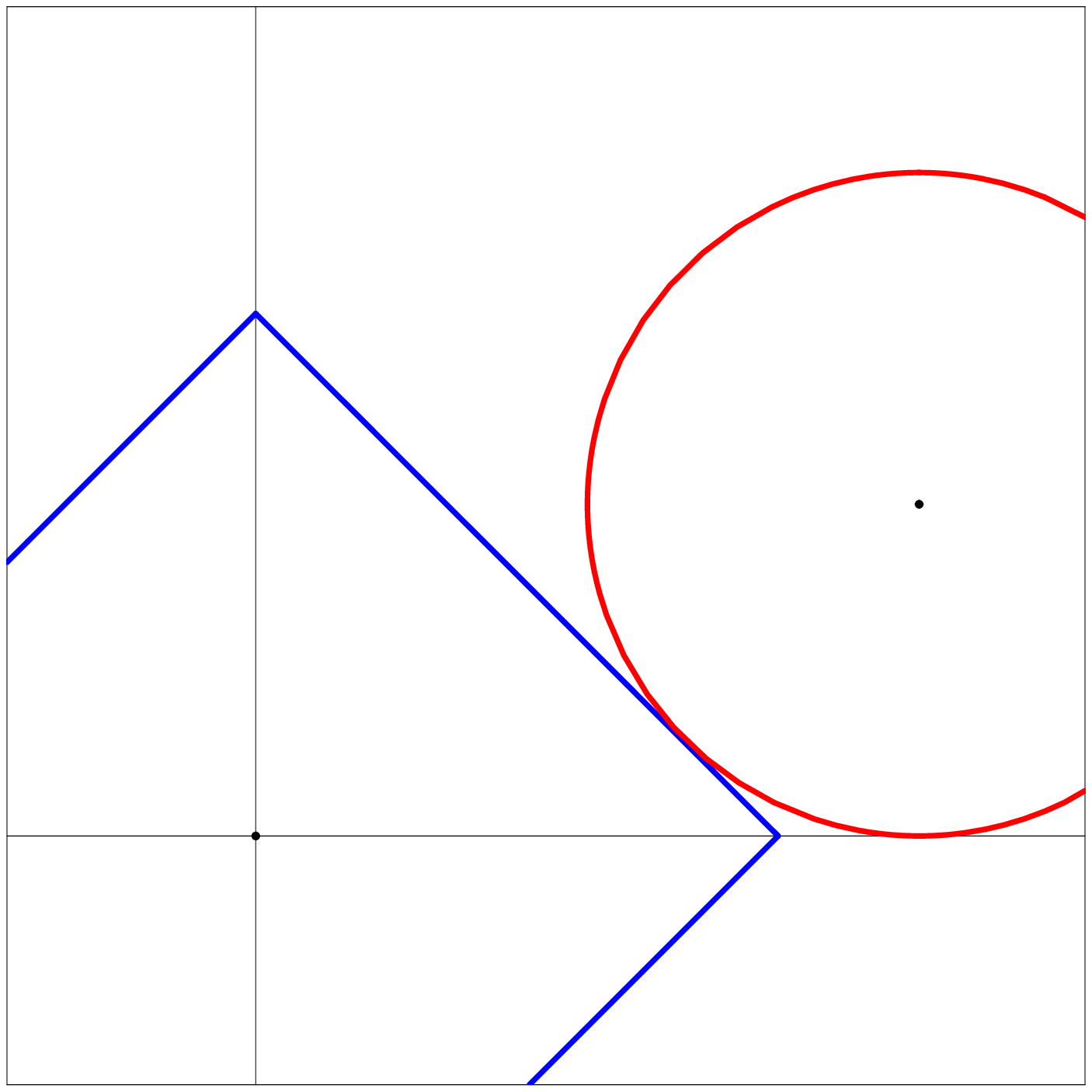}}\quad
\subfigure[Moderately sized $\ell_1$ penalty: tangency reaches
vertex.]{\includegraphics[scale = .2]{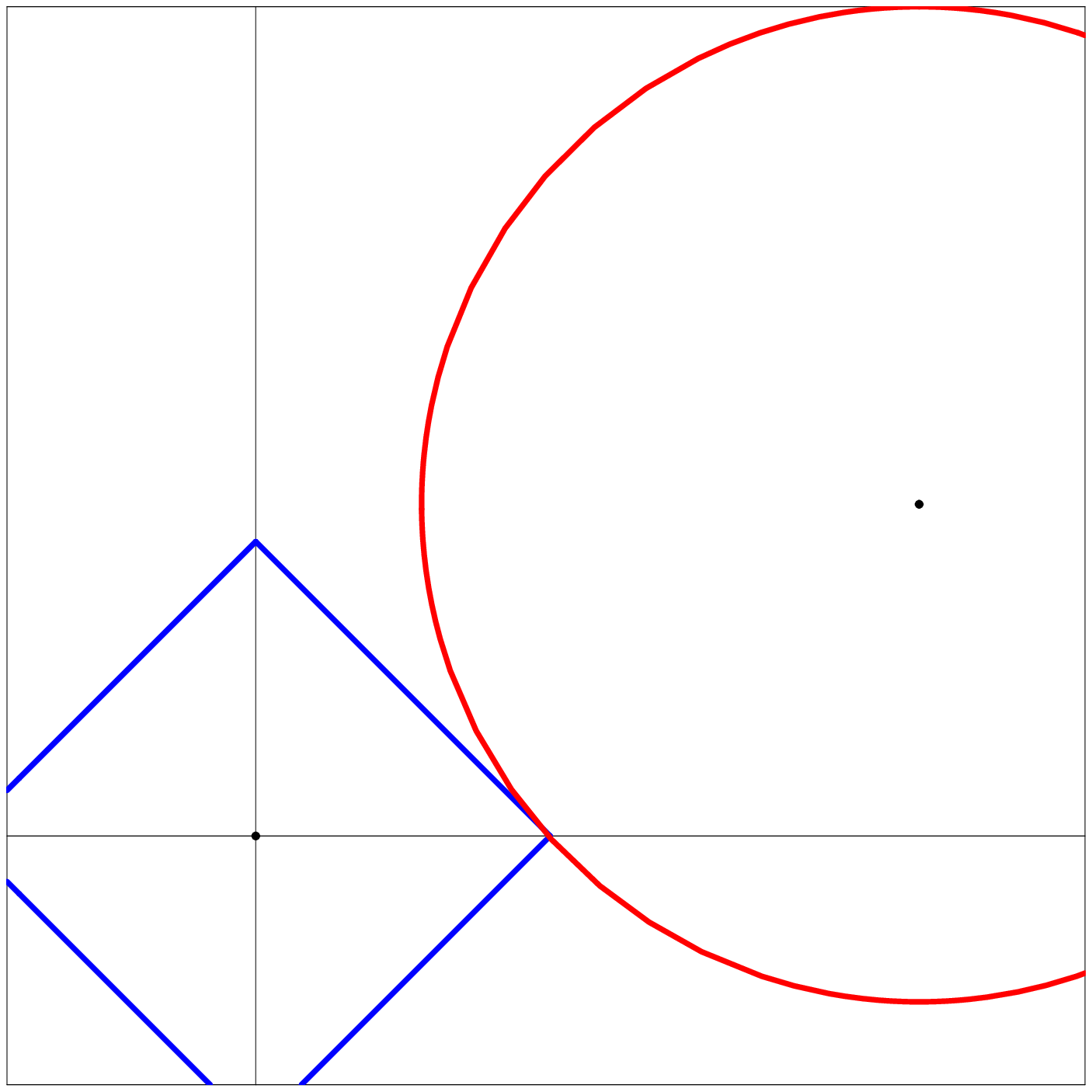}}\quad
\subfigure[Large $\ell_1$ penalty: tangency remains at vertex, moves
toward origin along axis.]{\includegraphics[scale =
.2]{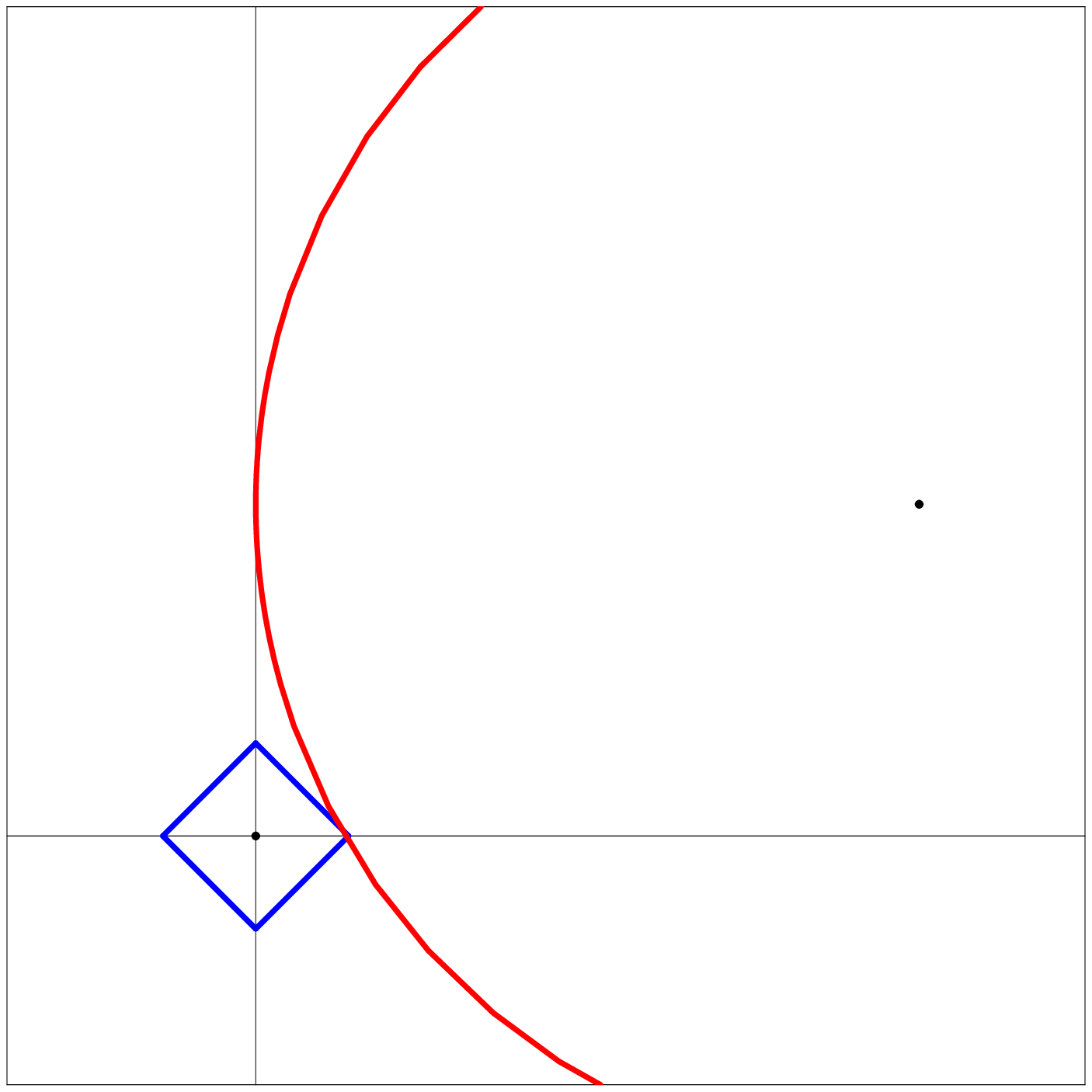}} \caption{\footnotesize $\ell_1$ penalties
promote sparsity: a geometric illustration in $2$ dimensions. In $N$
dimensions, the level sets of the $\ell_1$ norm are so-called cross
polytopes , while the level sets of the volatility term
(\ref{regression}) are typically ellipsoids. In 2 dimensions, we
illustrate these as diamonds and circles respectively.  The
minimizer of (\ref{lasso}) must be located at a point where the
level sets of the $\ell_1$ term and the volatility term are tangent.
There is a continuous family of such points, and the solution to a
particular minimization will be determined by the relative sizes of
the terms in the objective function (\ref{lasso}).  When the
$\ell_1$ norm is given a small weight in the objective function,
optimal solutions lie on level sets of large $\ell_1$ norm.  These
large $\ell_1$ level sets intersect the volatility level sets at
generic points where many, if not all, of the entries in $\w$ differ
from zero. However, as the weight on the $\ell_1$ term is increased,
the tangency moves onto smaller level sets of the $\ell_1$ norm.
Since the $\ell_1$ ball is ``pointy," the tangencies move toward the
corners of the $\ell_1$ level sets where more and more entries in
$\w$ are equal to $0$. Moreover, as Figure \ref{fig:ell-p} shows,
any $\ell_p$ penalty with $0\leqslant p\leqslant 1$ will lead to
``pointy" level sets that give rise to a similar sparsification
effect.} \label{fig:sparsity}
\end{figure}

\item It regulates the amount of {\em shorting} in the
portfolio designed by the optimization process. Because of the
constraint (\ref{total_weight_constraint}), an equivalent form of
the objective function in (\ref{lasso}) is
\begin{equation}
\|\rho \1_T - \pmb{R} \w\|^2_2+ 2\tau\!\!\!\sum_{i \mbox{ with
}w_i <0}|w_i|\,+\, \tau, \label{lasso2}
\end{equation}
in which the last term is of course irrelevant for the
optimization process. Under the constraint
(\ref{total_weight_constraint}), the $\ell_1$ penalty is thus
equivalent to a penalty on short positions.

The no-short-positions optimal portfolio, obtained by minimizing
(\ref{regression}) under the {\em three} constraints given by not
only (\ref{total_return_constraint}) and
(\ref{total_weight_constraint}) but also the additional restriction
$w_i \geqslant 0$ for $i=1,\ldots,N$, is in fact the optimal
portfolio for (\ref{lasso2}) in the limit of extremely large values
of $\tau$. As the high $\tau$ limit of our framework, it is
completely natural that the positive solution should be quite
sparse; this sparsity of optimal no-short-positions portfolios can
indeed also be observed in practice. (See Section 4.) We note that
the literature has focused on the stability of positive solutions,
but seems to have overlooked the sparsity of such solutions. This
may be due, possibly, to the use of iterative numerical optimization
algorithms and a stopping criterion that halted the optimization
before most of the components had converged to their zero limit. By
decreasing $\tau$ in the $\ell_1$-penalized objective function to be
optimized, one relaxes the constraint without removing it
completely; it then no longer imposes positivity absolutely, but
still penalizes overly large negative weights.

\item It {\em stabilizes} the problem. By imposing a penalty on
the size of the coefficients of $\w$ in an appropriate way, we
reduce the sensitivity of the optimization to the possible
collinearities between the assets. In \citet{Dau04}, it is proved
(for the unconstrained case) that any $\ell_p$ penalty on $\w$, with
$1\leqslant p \leqslant 2$, suffices to stabilize the minimization
of (\ref{regression}) by regularizing the inverse problem. The
stability induced by the $\ell_1$ penalization is extremely
important; indeed, it is such stability property that makes
practical, empirical work possible with only limited training data.
For example, \citet{DGR} show that this regularization method can be
used to produce accurate macroeconomic forecasts using many
predictors. Figure \ref{fig:ell-p} shows the geometric impact of an
$\ell_p$ penalty for the unconstrained problem for various values of
$p$.  It also illustrates why only $\ell_p$ penalties with $0
\leqslant p \leqslant 1$ are able to promote sparsity.

\begin{figure}
\centering \includegraphics[scale=.2]{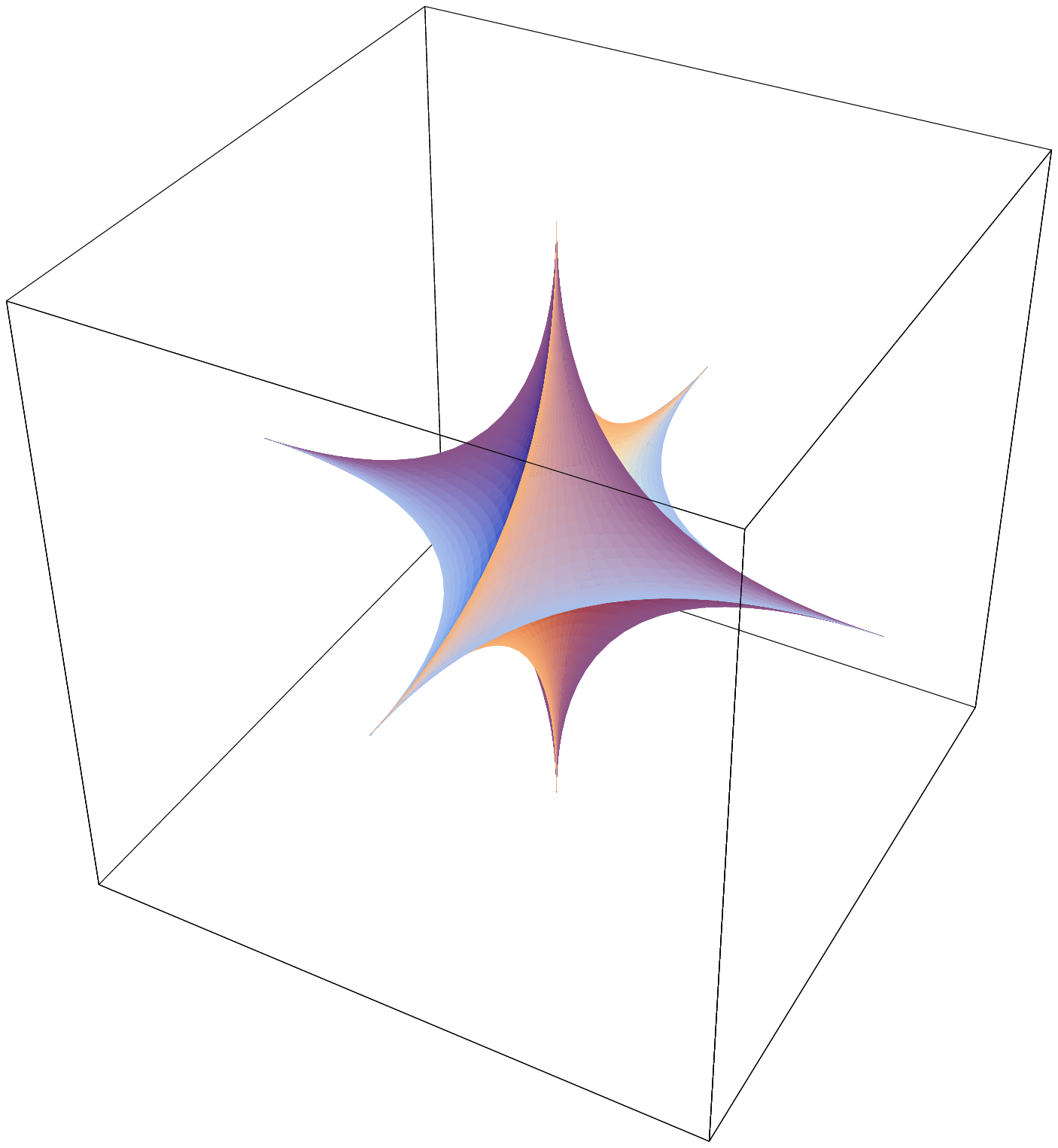}\quad
\includegraphics[scale=.2]{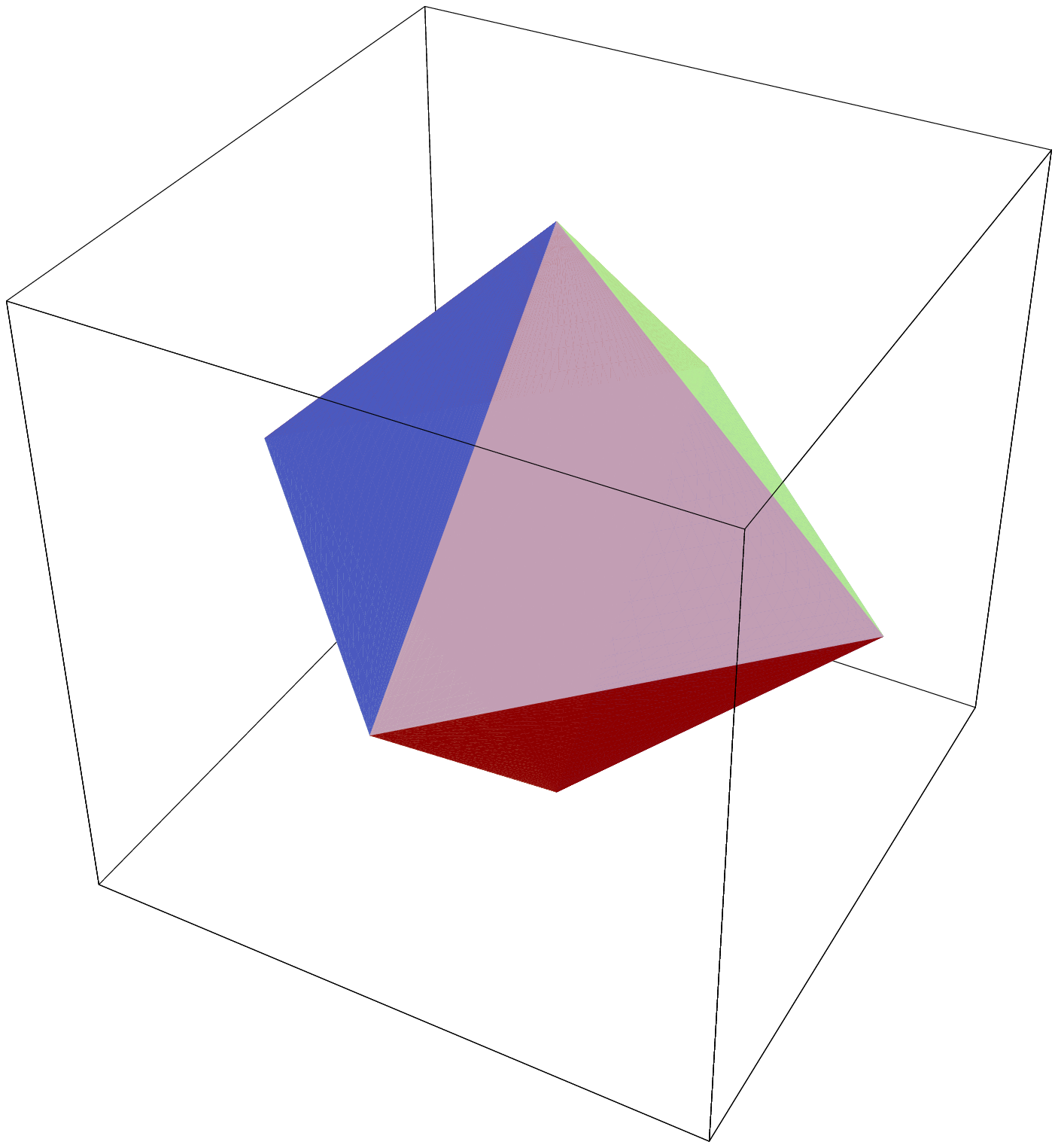}\quad \includegraphics[scale=.2]{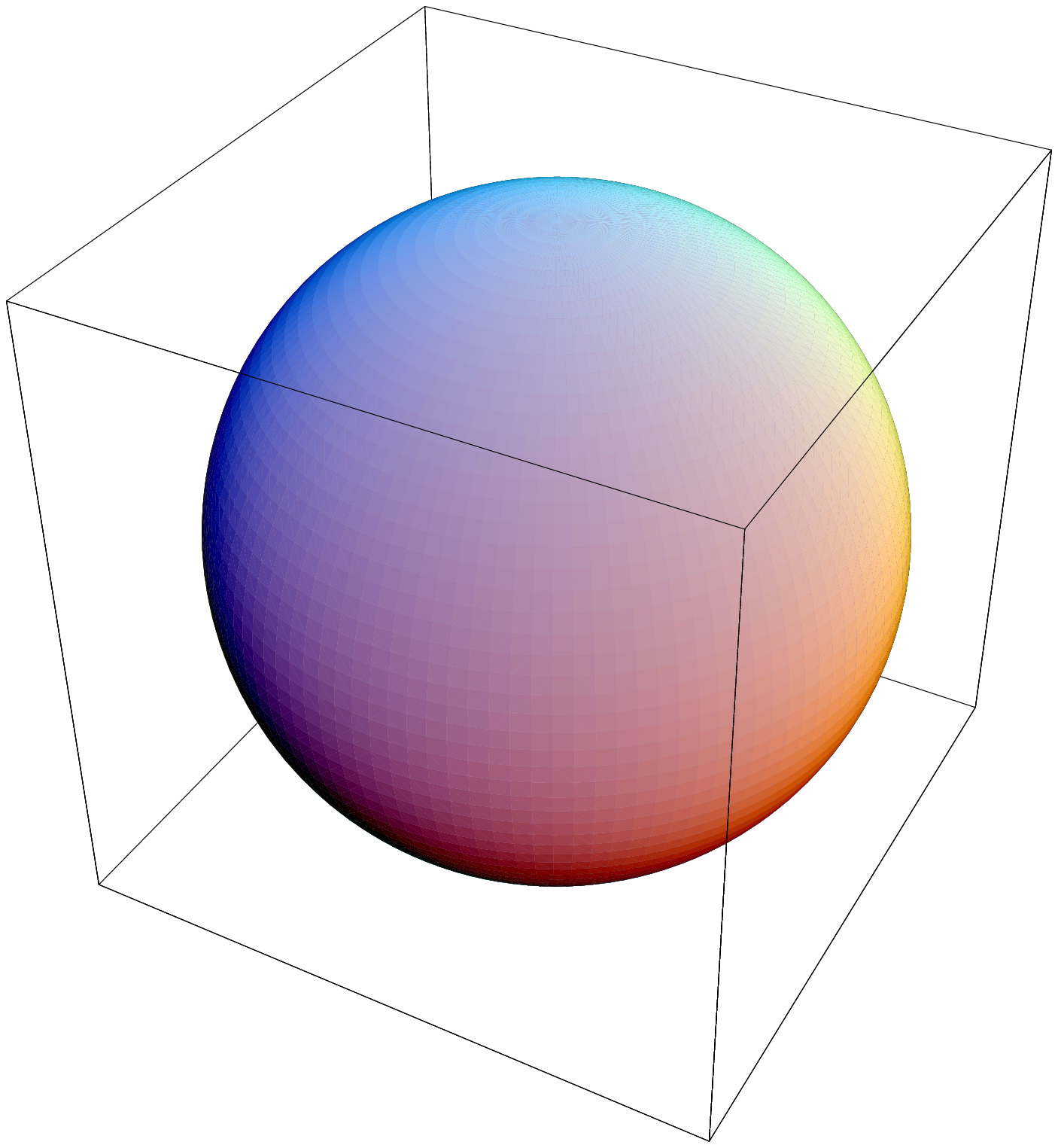}\quad
\includegraphics[scale=.2]{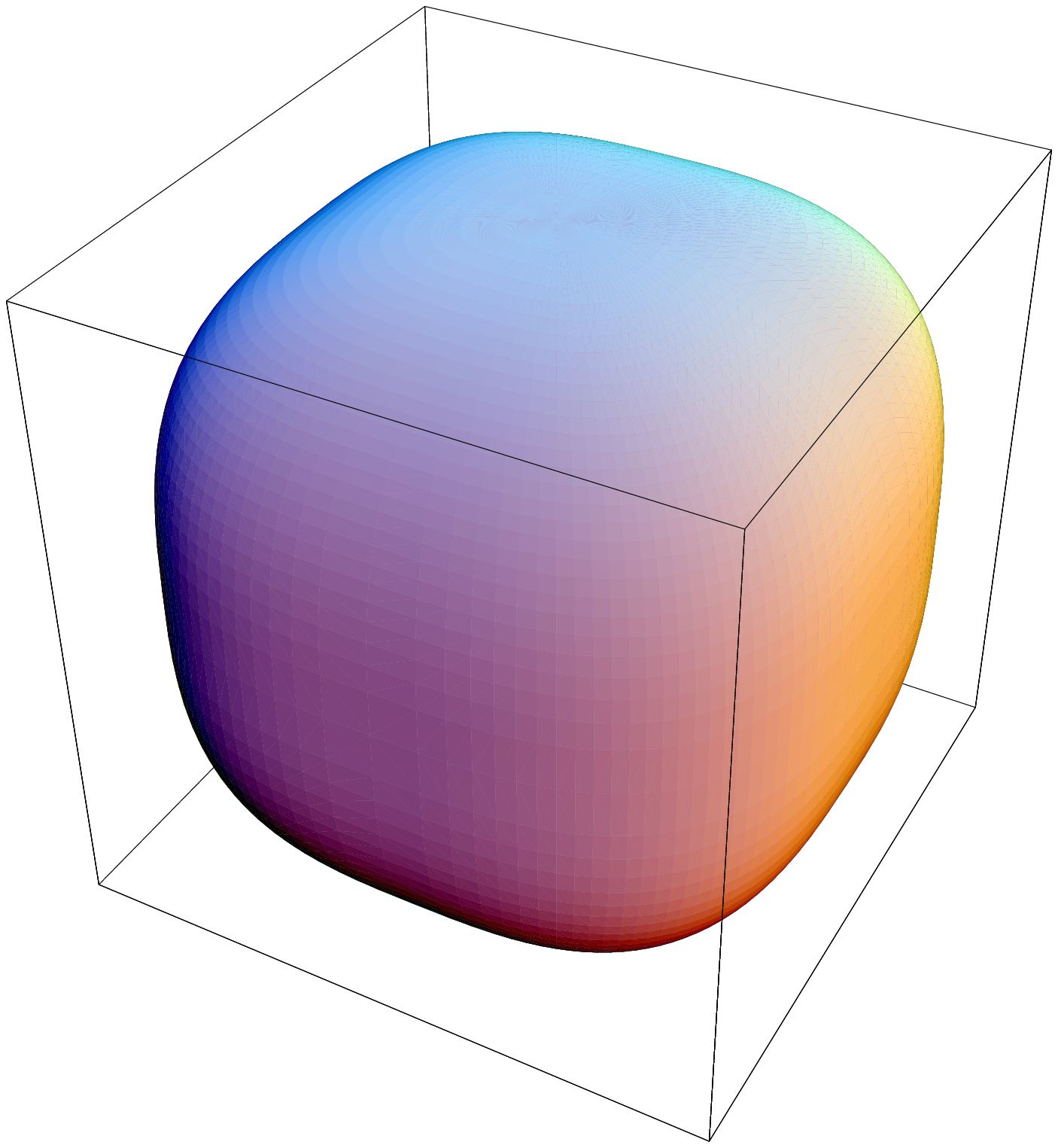} \caption{\footnotesize A
geometric look at regularization penalties. The panels above depict
sets of fixed $\ell_p$ norm for various values of $p$ (from left to
right, $p=1/2,1,2,3$), in 3 dimensions. For any fixed $p$, adding an
$\ell_p$ penalty to a given minimization encourages solutions to
stay within regions around the origin defined by scaled versions of
the $\ell_p$ ball. For $p \geqslant 1$ the amended minimization
problem remains convex, and thus algorithmically more tractable; on
the other hand $\ell_p$ penalties with $0 \leqslant p \leqslant 1$
encourage sparsity (see Figure \ref{fig:sparsity}). In our
optimization, we focus on the case $p = 1$, which has both desirable
features.} \label{fig:ell-p}
\end{figure}

\item It makes it possible to account for {\em transaction costs}
in a natural way. In addition to the choice of the securities they
trade, real-world investors must also concern themselves with the
transaction costs they will incur when acquiring and liquidating the
positions they select. Transaction costs in a liquid market can be
modeled by a two-component structure: one that is a fixed
``overhead'', independent of the size of the transaction, and a
second one, given by multiplying the transacted amount with the
market-maker's bid-ask spread applicable to the size of the
transaction.

For large investors, the overhead portion can be neglected; in that
context, the total transaction cost paid is just $\sum_{i=1}^N s_i
\,|w_i| $, the sum of the products of the absolute trading volumes
$|w_i|$ and bid-ask spreads $s_i$ for the securities $i=1, \ldots,
N$.  We assume that the bid-ask spread is the same for all assets
and constant for a wide range of transaction sizes.  In that case,
the transaction cost is then effectively captured by an $\ell_1$
penalty\footnote{Our methodology can be easily generalized to
asset-dependent bid-ask spreads --- see Section 5.}.

For small investors, the overhead portion of the transaction costs
is non negligible. In the case of a very small investor, this
portion may even be the only one worth considering; if the
transaction costs are asset-independent, then the total cost is
simply proportional to the number $K$ of assets selected (i.e.
corresponding to non-zero weights), a number sometimes referred to
as $ \| \w\|_0$, the $\ell_0$ norm of the weight vector. Like an
$\ell_1$ sum, this $\ell_0$ sum can be incorporated into the
objective function to be minimized; however, $\ell_0$-penalized
optimization is computationally intractable when more than a handful
of variables are involved because its complexity is essentially
combinatorial in nature, and grows super-exponentially with the
number of variables. For this reason, one often replaces the
$\ell_0$ penalty, when it occurs, by its much more tractable
(convex) $\ell_1$-penalty cousin, which has similar
sparsity-promoting properties. In this sense, our $\ell_1$
penalization is thus ``natural'' even for small investors.

\end{itemize}

\section{Optimization strategy}

We first quickly review the unconstrained case, i.e. the
minimization of the objective function (\ref{lasso}), and then
discuss how to deal with the constraints
(\ref{total_return_constraint}) and (\ref{total_weight_constraint}).

Various algorithms can be used to solve (\ref{lasso}). For the
values of the parameters encountered in the portfolio construction
problem, a particularly convenient algorithm is given by the
homotopy method \citep{osb00a, osb00b}, also known as {\bf L}east
{\bf A}ngle {\bf R}egression or LARS \citep{Efr04}. This algorithm
seeks to solve (\ref{lasso}) for a range of values of $\tau$,
starting from a very large value, and gradually decreasing $\tau$
until the desired value is attained. As $\tau$ evolves, the optimal
solution $\w^{[\tau]}$ moves through $\mathbb{R}^N$, on a piecewise
affine path. As such, to find the whole locus of solutions for
$\w^{[\tau]}$ we need only find the critical points where the slope
changes. These slopes are thus the only quantities that need to be
computed explicitly, besides the breakpoints of the piecewise linear
(vector-valued) function. For every value of $\tau$,  the entries
$j$ for which $w_j \,\neq \,0$, are said to constitute the {\em
active set} $\mathcal{A}_{\tau}$. Typically, the number of elements
of $\mathcal{A}_{\tau}$ increases as $\tau$ decreases. However, this
is not always the case: at some breakpoints entries may need to be
removed from $\mathcal{A}_{\tau}$; see e.g. \citep{Efr04}.

When the desired minimizer contains only a small number $K$ of
non-zero entries, this method is very fast. At each breakpoint, the
procedure involves solving a linear system of $k$ equations with $k$
unknowns, $k$ being the number of active variables, which increases
until $K$ is reached. This imposes a pragmatic feasibility upper
bound on $K$.

The homotopy/LARS algorithm applies to the {\em unconstrained}
$\ell_1$-penalized regression. The problem of interest to us,
however, is the minimization of (\ref{lasso}) {\em under the
constraints} (\ref{total_return_constraint},
\ref{total_weight_constraint}), in which case the original algorithm
does not apply. In the Appendix, we show how the homotopy/LARS
algorithm can be adapted to deal with a general $\ell_1$-penalized
minimization problem with linear constraints, allowing us to find:
\begin{equation}
\w^{[\tau]}  =  \arg \min_{\w \in H} \left[ \|\y-\R\w\|^2_2+ \tau
\|\w\|_1\right]  \label{gen_funct_w_constr}
\end{equation}
where $H$ is a prescribed affine subspace, defined by the linear
constraints. The adapted algorithm consists again of starting with
large values of $\tau$, and shrinking $\tau$ gradually until the
desired value is reached, monitoring the solution, which is still
piecewise linear, and solving a linear system at every breakpoint in
$\tau$. Because of the constraints, the initial solution (for large
values of $\tau$) is now more complex (in the unconstrained case, it
is simply equal to zero); in addition, extra variables (Lagrange
multipliers) have to be introduced that are likewise piecewise
linear in $\tau$, and the slopes of which have to be recomputed at
every breakpoint.

In the particular case of the minimization (\ref{lasso}) under the
constraints (\ref{total_return_constraint}),
(\ref{total_weight_constraint}), an interesting interplay takes
place between (\ref{total_weight_constraint}) and the
$\ell_1$-penalty term. When the weights $w_i$ are all non-negative,
the constraint (\ref{total_weight_constraint}) is equivalent to
setting $\|\mathbf{w}\|_1\,=\,1$. Given that the $\ell_1-$penalty
term takes on a fixed value in this case, minimizing the quadratic
term only (as in (\ref{regression})) is thus equivalent to
minimizing the penalized objective function in (\ref{lasso}), {\em
for non-negative weights} $w_i$. This is consistent with the
observation made by \citet{JaMa03} that a restriction to
non-negative-weights-only can have a regularizing effect on
Markowitz's portfolio construction.

The following mathematical observations have interesting
consequences. Suppose that the two weight vectors $\w^{[\tau_1]}$
and $\w^{[\tau_2]}$ are minimizers of (\ref{lasso}), corresponding
to the values $\tau_1$ and $\tau_2$ respectively, and both satisfy
the two constraints (\ref{total_return_constraint}),
(\ref{total_weight_constraint}). By using the respective
minimization properties of $\w^{[\tau_1]}$ and $\w^{[\tau_2]}$, we
obtain
\begin{eqnarray}
\|\rho \1_T - \pmb{R} \w^{[\tau_1]}\|^2_2\,+\,
\tau_1\|\w^{[\tau_1]}\|_1 & \leqslant &
\|\rho \1_T - \pmb{R} \w^{[\tau_2]}\|^2_2\,+\,
\tau_1\|\w^{[\tau_2]}\|_1 \nonumber\\
& = & \|\rho \1_T - \pmb{R} \w^{[\tau_2]}\|^2_2\,+\,
\tau_2\|\w^{[\tau_2]}\|_1 \nonumber\\
 & & \qquad\qquad\,+\, (\,\tau_1 - \tau_2\,)\,
\|\w^{[\tau_2]}\|_1 \nonumber\\
& \leqslant & \|\rho \1_T - \pmb{R} \w^{[\tau_1]}\|^2_2\,+\,
\tau_2\|\w^{[\tau_1]}\|_1 \nonumber\\
 & & \qquad\qquad\,+\, (\,\tau_1 - \tau_2\,)\,
\|\w^{[\tau_2]}\|_1 \nonumber\\
& = & \|\rho \1_T - \pmb{R} \w^{[\tau_1]}\|^2_2\,+\,
\tau_1\|\w^{[\tau_1]}\|_1 \nonumber\\
& & \,+\, (\,\tau_1 - \tau_2\,)\, \left(\,\|\w^{[\tau_2]}\|_1 -
\|\w^{[\tau_1]}\|_1\,\right),\nonumber 
\end{eqnarray}
which implies that
\begin{equation}
(\,\tau_1 \,-\,  \tau_2\,)\, \left(\,\|\w^{[\tau_2]}\|_1 \,-\,
\|\w^{[\tau_1]}\|_1\,\right) \,\geqslant \, 0. \label{chain2}
\end{equation}
If some of the $w_i^{[\tau_2]}$ are negative, but all the entries in
$\w^{[\tau_1]}$ are positive or zero, then we have
$\|\w^{[\tau_2]}\|_1 \,>\, \left| \sum_{i=1}^N w_i^{[\tau_2]}\right|
\, = \, 1$; on the other hand, $\|\w^{[\tau_1]}\|_1\,=\,1$ (because
the $w_i^{[\tau_1]}$ are all non-negative),  implying
$\|\w^{[\tau_2]}\|_1 \,>\, \|\w^{[\tau_1]}\|_1$ . In view of
(\ref{chain2}), this means that $\tau_1 \,\geqslant\, \tau_2$.

It follows that the optimal portfolio with non-negative entries
obtained by our minimization procedure corresponds to the {\em
largest} values of $\tau$, and thus typically to the {\em
sparsest} solution (since the penalty term, promoting sparsity, is
weighted more heavily). This particular portfolio is a minimizer
of (\ref{lasso}), under the constraints
(\ref{total_return_constraint}) and
(\ref{total_weight_constraint}), for all $\tau$ larger than some
critical value $\tau_0$. For smaller $\tau$ the optimal portfolio
will contain at least one negative weight and will typically
become less sparse. However, as in the unconstrained case, this
need not happen in a monotone fashion.

Although other optimization methods could be used to compute the
sparse portfolios we define, the motivation behind our choice of a
constrained homotopy/LARS algorithm is the fact that we are only
interested in computing portfolios involving only a small number of
securities and that we use the parameter $\tau$ to tune this number.
Whereas other algorithms would require separate computations to find
solutions for each value of $\tau$, a particularly nice feature of
our LARS-based algorithm is that, by exploiting the piecewise linear
dependence of the solution on $\tau$, it obtains, in one run, the
weight vectors for all values of $\tau$ (i.e. for all numbers of
selected assets) in a prescribed range.

\section{Empirical application}

In this section we apply the methodology described above to
construct optimal portfolios and evaluate their out-of-sample
performance.

We present two examples, each of which uses a universe of
investments compiled by Fama and French\footnote{These data are
available from the site\\
\url{http://mba.tuck.dartmouth.edu/pages/faculty/ken.french/data_library.html},
to which we refer for more details on these portfolios.}.  In the
first example, we use 48 industry sector portfolios (abbreviated to
FF48 in the remainder of this paper). In the second example, we use
100 portfolios formed on size and book-to-market
(FF100).\footnote{These portfolios are the intersections of 10
portfolios formed on size and 10 portfolios formed on the ratio of
book equity to market equity.} In both FF48 and FF100, the
portfolios are constructed at the end of each June.

\subsection{Example 1: FF48}

Using the notation of Section 2, $r_{i,t}$ is the annualized return
in month $t$ of industry $i$, where $i = 1, \ldots, 48$. We evaluate
our methodology by looking at the out-of-sample performances of our
portfolios during the last 30 years in a simulated investment
exercise.

\begin{figure}
\centering \includegraphics[scale=0.3]{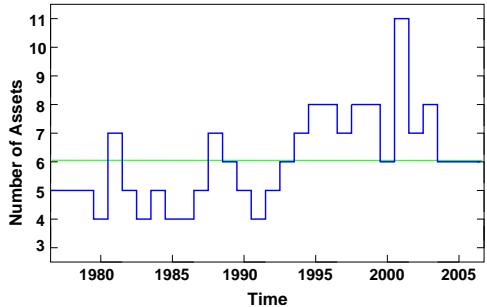}
\caption{\footnotesize Number of assets without short positions, for
FF48. The number of active assets $K_{\mbox{\tiny{pos.}}}$ in
$\w^{\mbox{\tiny{pos.}}}$, the optimal portfolio without short
positions, from year to year. This number varies from 4 to 11; the
average over 30 years is around 6.}\label{optimal_number}
\end{figure}

Each year from 1976 to 2006, we construct a collection of optimal
portfolios by solving an ensemble of minimizations of
(\ref{lasso}) with constraints (\ref{total_return_constraint}) and
(\ref{total_weight_constraint}). For each time period, we carry
out our optimization for a sufficiently wide range of $\tau$ so as
to produce an ensemble of portfolios containing different numbers
of active positions; ideally we would like to construct portfolios with $K$ securities, for all
values of $K$ between 2 and 48. As explained below, we do not
always obtain all the low values of $K$; typically we find
optimal portfolios only for $K$ exceeding a minimal value
$K_{\mbox{\tiny{min}}}$, that varies from year to year. (See
Figure \ref{optimal_number}).  To estimate the necessary return
and covariance parameters, we use data from the preceding 5 years
(60 months). At the time of each portfolio construction, we set
the target return, $\rho$, to be the average return achieved by
the na\"{i}ve, evenly-weighted portfolio over the previous 5
years.

For example, our first portfolio construction takes place at
the end of June 1976. To determine $\pmb{R}$ and $\widehat{\mmu}$,
we use the historical returns from July 1971 until June 1976. We
then solve the optimization problem using this matrix and vector,
targeting an annualized return of $6.60\%$ ($\rho=0.066$),
equal to
the average historical return, from July 1971 until June 1976,
obtained by a portfolio in which all industry sectors are given
the equal weight 1/48. We compute the weights of optimal
solutions $\w^{[\tau]}$ for $\tau$ ranging from large to small
values. We select these portfolios according to some criterion we
would like to meet. We could, e.g. target a fixed total number of
active positions, or limit the number of short positions; see
below for examples. Once a portfolio is thus fixed, it is kept
from July 1976 until June 1977, and its returns are recorded. At
the end of June 1977, we repeat the same process, using training
data from July 1972 to June 1977 to compute the composition of a
new collection of portfolios. These portfolios are observed from
July 1977 until June 1978 and their returns are recorded. The same
exercise is repeated every year with the last ensemble of
portfolios constructed at the end of June 2005.

Once constructed, the portfolios are  thus held through June of the
next year and their monthly  out-of-sample returns are observed.
These monthly returns, for all the observation years together,
constitute a time series; for a given period (whether it is the full
1976$\rightarrow$2006 period, or sub-periods), all the monthly
returns corresponding to this period are used to compute the average
monthly return $m$, its standard deviation $\sigma$, and their ratio
$m /\sigma$, which is then the Sharpe ratio measuring the trade-off,
corresponding to the period, between returns and volatility of the
constructed portfolios.

We emphasize that the {\em sole purpose} of carrying out
the portfolio construction multiple times, in successive years, is
to collect data from which to evaluate the effectiveness of the
portfolio construction strategy. These constructions from scratch
in consecutive years are {\em not} meant to model the behavior of
a single investor; they model, rather, the results obtained by
{\em different} investors who would follow the same strategy to
build their portfolio, starting in different years. A single
investor might construct a starting portfolio according to the
strategy described here, but might then, in subsequent years,
adopt a sparse portfolio adjustment strategy such as described in
Section 5.

We compare the performance of our strategy to that of a benchmark
strategy comprising an equal investment in each available
security. This $1/N$ strategy is a tough benchmark since it has
been shown to outperform a host of optimal portfolio strategies
constructed with existing optimization procedures \citep{DeM07}.
To evaluate the $1/N$ strategy portfolios for the
FF48 assets, we likewise observe the monthly returns for a certain
period (a 5-year break-out period or the full 30-year period), and
use them to compute the average mean return $m$, the standard
deviation $\sigma$, and the Sharpe ratio $m/\sigma$.

We carried out the full procedure following several possible
guidelines. The first such guideline is to pick the optimal
portfolio $\w^{\mbox{\tiny{pos.}}}$ that has only non negative
weights $w_i$, i.e. the optimal portfolio without short positions.
As shown in Section 3,  this portfolio corresponds to the largest
values of the penalization constant $\tau$; it typically is also the
optimal portfolio with the fewest assets. Figure
\ref{optimal_number} reports the number of active assets of this
optimal no-short-positions portfolio from year to year. This number
varies from a minimum of  4 to a maximum of 11, with an average of
around 6; note that this is quite sparse in a 48 asset universe.
Table \ref{table1} reports statistics to evaluate the performances
of the optimal no-short-positions portfolio. We give the statistics
for the whole sample period as well as for consecutive sub-periods
extending over 5 years each, comparing these to the portfolio that
gives equal weight to the 48 assets. The table shows that the
optimal no-short-positions portfolio significantly outperforms the
benchmark both in terms of returns and in terms of volatility; this
result holds for the full sample period as well as for the
sub-periods. Note that most of the gain comes from the smaller
variance of the sparse portfolio around its target return, $\rho$.

A second possible guideline for  selecting the portfolio
construction strategy is  to  target a particular number of assets,
or a particular narrow range for this number. For instance, users
could decide to pick, every year, the optimal portfolio that has
always more than 8 but at most 16 assets. Or the investor may decide
to select an optimal portfolio with, say, exactly 13 assets. In this
case, we would carry out the minimization, decreasing $\tau$ until
we reach the breakpoint value where the number of assets in the
portfolio reaches 13. We shall denote the corresponding weight
vector by $\w^{\mbox{\tiny{13}}}$.

\begin{table}
\caption{Performance of the sparse portfolio with no short-selling,
for FF48.}

\centering
\begin{tabular}{|c||c|c|c||c|c|c|}
\multicolumn{7}{c}{}\\
\hline Evaluation&\multicolumn{3}{c||}{$w_i \geqslant 0$}&
\multicolumn{3}{c|}{Equal} \\
period&\multicolumn{3}{c||}{for all $i$}&\multicolumn{3}{c|}{weight} \\
&$m$&$\sigma$&$S$&$m$&$\sigma$&$S$\\
\hline \hline
06/76-06/06 &   17 & 41 & 41 &   17 & 61 & 27\\
\hline
07/76-06/81  &  23 & 48 & 49 &   29 & 66 & 44\\
07/81-06/86  &  23 & 41 & 57 &   18 & 58 & 31\\
07/86-06/91  &   9 & 45 & 20 &    5 & 72 &  7\\
07/91-06/96  &  16 & 26 & 62 &   18 & 41 & 44\\
07/96-06/01  &  16 & 40 & 40 &   11 & 67 & 17\\
07/01-06/06  &  13 & 43 & 30 &   18 & 60 & 30 \\\hline
\end{tabular}
\parbox{9cm}{\footnotesize Monthly mean return $m$, standard
deviation of monthly return $\sigma$, and corresponding Sharpe ratio
$S$ (expressed in \%) for the optimal no-short-positions portfolio,
as well as for the $1/N$-strategy portfolio. Both portfolio
strategies are tested for their performance over twelve consecutive
months immediately following their construction; their returns are
pooled over several years to compute $m$, $\sigma$ and $S$, as
described in the text. Statistics reflect the performance an
investor would have achieved, on average, by constructing the
portfolio on July 1 one year, and keeping it for the next twelve
months, until June 30 of the next year, with the average taken over
5 years for each of the break-out periods, over 30 years for the
full period. }

 \label{table1}
\end{table}

\begin{figure}\centering
\includegraphics[scale=0.5]{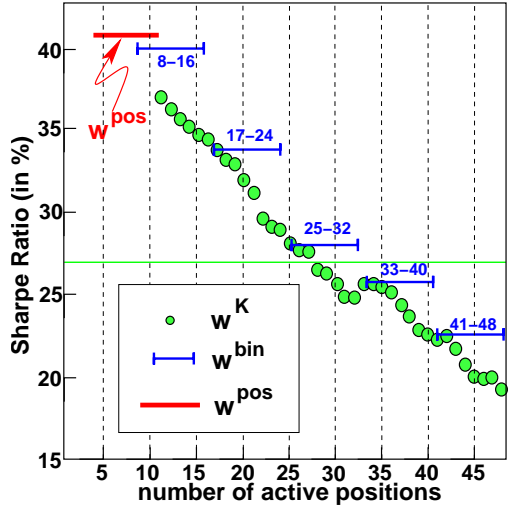}

\caption{\footnotesize Outperforming the $1/N$ strategy, for FF48.
The Sharpe ratio, for the full period 1976-2006, for several
portfolios: $\w^{\mbox{\tiny{pos.}}}$, the optimal portfolio without
short positions (red); the portfolios $\w^K$ with a fixed number,
$K$, of active positions, where $K$ ranges from 11 to 48 (green
dots). $\w^{\mbox{\tiny{pos.}}}$ is indicated by a fat solid
horizontal red bar, stretching from 4 to 11 (its minimum to maximum
number of assets; see also Figure \ref{optimal_number}.) The binned
portfolios $\w^{\mbox{\tiny{bin}}}$ are marked by horizontal blue
bars indicating the range of the corresponding bin. The horizontal
green line indicates the Sharpe ratio for the $1/N$ portfolio, in
which equal amounts are invested in all the assets; a large number
of our portfolios significantly outperform this benchmark, with only
the least sparse (most weakly penalized) choices underperforming it.
}\label{sharpe}
\end{figure}

\begin{table}
\caption{Empirical results for FF48.}

\centering \resizebox{\textwidth}{!}{
\begin{tabular}{|c|c|c|c|c|c|c|c|c|c|c|c|c|c|c|c|c|c|c|c|c|c|}
\hline
Evaluation&\multicolumn{3}{c|}{bin=8-16}&\multicolumn{3}{c|}{bin=17-24}&\multicolumn{3}{c|
}{bin=25-32}&\multicolumn{3}{c|}{bin=33-40}&\multicolumn{3}{c|}{bin=41-48}&\multicolumn{3}{c|}{$w_i\geqslant 0$} \\
period&$m$&$\sigma$&$S$&m&$\sigma$&$S$&$m$&$\sigma$&$S$&$m$&$\sigma$&$S$&$m$&$\sigma$&$S$&$m$&$\sigma$&$S$\\
\hline
\!\!07/76-06/06\!\!&\!\!16\!\!&\!\!40\!\!&\!\!40\!\!&\!\!14\!\!&\!\!40\!\!&\!\!34\!\!&\!\!12\!\!&\!\!43\!\!&\!\!28\!\!&\!\!12\!\!&\!\!47\!\!&\!\!26\!\!&\!\!12\!\!&\!\!54\!\!&\!\!22\!\!&\!\!17\!\!&\!\!41\!\!&\!\!41\!\!\\
\hline
\!\!07/76-06/81\!\!&\!\!20\!\!&\!\!41\!\!&\!\!50\!\!&\!\!18\!\!&\!\!39\!\!&\!\!46\!\!&\!\!19\!\!&\!\!40\!\!&\!\!49\!\!&\!\!22\!\!&\!\!43\!\!&\!\!50\!\!&\!\!23\!\!&\!\!50\!\!&\!\!46\!\!&\!\!23\!\!&\!\!48\!\!&\!\!49\!\!\\
\!\!07/81-06/86\!\!&\!\!25\!\!&\!\!42\!\!&\!\!58\!\!&\!\!23\!\!&\!\!44\!\!&\!\!52\!\!&\!\!24\!\!&\!\!46\!\!&\!\!52\!\!&\!\!23\!\!&\!\!50\!\!&\!\!46\!\!&\!\!22\!\!&\!\!56\!\!&\!\!39\!\!&\!\!23\!\!&\!\!41\!\!&\!\!57\!\!\\
\!\!07/86-06/91\!\!&\!\!8\!\!&\!\!43\!\!&\!\!18\!\!&\!\!7\!\!&\!\!44\!\!&\!\!15\!\!&\!\!4\!\!&\!\!47\!\!&\!\!9\!\!&\!\!4\!\!&\!\!51\!\!&\!\!7\!\!&\!\!5\!\!&\!\!57\!\!&\!\!8\!\!&\!\!9\!\!&\!\!45\!\!&\!\!20\!\!\\
\!\!07/91-06/96\!\!&\!\!15\!\!&\!\!26\!\!&\!\!57\!\!&\!\!13\!\!&\!\!27\!\!&\!\!47\!\!&\!\!12\!\!&\!\!33\!\!&\!\!36\!\!&\!\!12\!\!&\!\!41\!\!&\!\!30\!\!&\!\!11\!\!&\!\!52\!\!&\!\!21\!\!&\!\!16\!\!&\!\!26\!\!&\!\!62\!\!\\
\!\!07/96-06/01\!\!&\!\!16\!\!&\!\!41\!\!&\!\!38\!\!&\!\!9\!\!&\!\!42\!\!&\!\!22\!\!&\!\!3\!\!&\!\!50\!\!&\!\!6\!\!&\!\!2\!\!&\!\!54\!\!&\!\!4\!\!&\!\!0\!\!&\!\!61\!\!&\!\!0\!\!&\!\!16\!\!&\!\!40\!\!&\!\!40\!\!\\
\!\!07/01-06/06\!\!&\!\!13\!\!&\!\!42\!\!&\!\!30\!\!&\!\!12\!\!&\!\!41\!\!&\!\!29\!\!&\!\!10\!\!&\!\!38\!\!&\!\!28\!\!&\!\!10\!\!&\!\!37\!\!&\!\!27\!\!&\!\!12\!\!&\!\!43\!\!&\!\!27\!\!&\!\!13\!\!&\!\!43\!\!&\!\!30\!\!\\
\hline
\end{tabular}}

\parbox{\textwidth}{\small Monthly mean return, $m$, standard deviation of monthly return,
$\sigma$, and corresponding monthly Sharpe ratio $S$ (expressed in
\%) for the optimal portfolios with  8-16, 17-24, 25-32, 33-40,
41-48 assets, as well as (again) the optimal no-short-positions
portfolio. Portfolios are constructed annually as described in the
text. Statistics reflect the performance an investor would have
achieved on average, by constructing the portfolio on July 1 one
year, and keeping it for the next twelve months, until June 30 of
the next year; the average is taken over several years: 5 for each
break-out period, 30 for the full period.}
\label{table2}\label{TableBins}
\end{table}

For a ``binned'' portfolio, such as the 8-to-16 asset portfolio,
targeting a narrow range rather than an exact value for the total
number of assets, we define the portfolio $\w^{\mbox{\tiny{8-16}}}$
by considering each year the portfolios $\w^K$ with $K$ between 8
and 16 (both extremes included), and selecting the one that
minimizes the objective function (\ref{regression}); if there are
several possibilities, the minimizer with smallest $\ell_1$ norm is
selected. The results are summarized in Figure \ref{sharpe}, which
shows the average monthly Sharpe ratio of different portfolios of
this type for the entire 30 year exercise. For several portfolio
sizes, we are able to significantly outperform the evenly-weighted
portfolio (the Sharpe ratio of which is indicated by the horizontal
line at 27\%). Detailed statistics are reported in Table
\ref{table2};  for comparison, Table \ref{table2} lists again the
results for the no-short-positions portfolio.

Notice that, according to Table \ref{table2}, the no-short-positions
portfolio outperforms all binned portfolios for the full 30 year
period; this is not systematically true for the break-out periods,
but even in those break-out periods where it fails to outperform all
binned portfolios, its performance is still close to that of the
best performing (and sparsest) of the binned portfolios. This
observation no longer holds for the portfolio constructions with
FF100, our second exercise --- see Figure \ref{sharpeFF100} below.

\subsection{Example 2: FF100}

\begin{figure}[b]\centering
\includegraphics[scale=0.3]{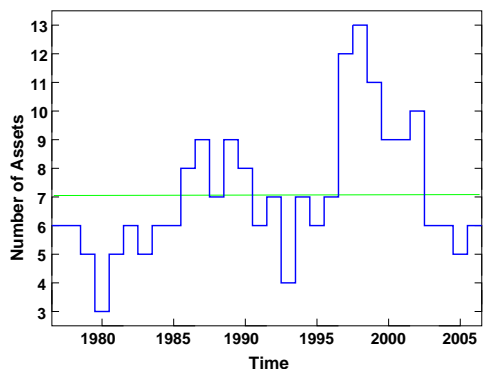}
\caption{\footnotesize Number of assets without short positions, for
FF100. The number of active assets $K_{\mbox{\tiny{pos.}}}$ in
$\w^{\mbox{\tiny{pos.}}}$, the optimal portfolio without short
positions, from year to year. Note that $K_{\mbox{\tiny{pos.}}}(t)$
is also the minimum value $K_{\mbox{\tiny{min}}}(t)$. For this data
set, our construction generates optimal portfolios for all values
from $K_{\mbox{\tiny{pos.}}}(t)$ to 60 in every year $t$.
$K_{\mbox{\tiny{pos.}}}(t)$  varies from 3 to 13; the average over
30 years is 7.} \label{Active_FF100}
\end{figure}

\begin{figure}\centering
\includegraphics[scale=0.5]{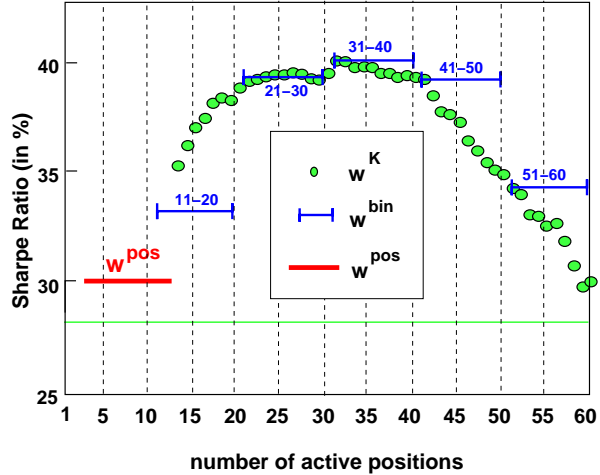}
\caption{\footnotesize Outperforming the $1/N$ strategy for FF100.
The Sharpe ratio, for the full period 1976-2006, for several
portfolios: $\w^{\mbox{\tiny{pos.}}}$, the optimal portfolio without
short positions (red); the ``binned'' portfolios $\w^{11-20}$,
$\w^{21-30}$, $\w^{31-40}$, $\w^{41-50}$ and $\w^{51-60}$ (blue);
and the portfolios $\w^K$ with a fixed number, $K$, of active
positions, where $K$ ranges from 13 to 60 (green dots circled in
black). $\w^{\mbox{\tiny{pos.}}}$ is indicated by a fat solid
horizontal red bar, stretching from 3 to 13 (its minimum to maximum
number of assets; see also Figure \ref{Active_FF100}.) In this
example, optimal sparse portfolios that allow short positions
significantly outperform both the evenly-weighted portfolio and the
optimal no-short-positions portfolio.}\label{sharpeFF100}
\end{figure}

Except for using a different collection of assets, this exercise is
identical in its methodology to what was done for FF48, so that we
do not repeat the full details here. Tables \ref{table3},
\ref{table4} and Figures \ref{Active_FF100}, \ref{sharpeFF100}
summarize the results; they correspond, respectively, to the results
given in Tables \ref{table1}, \ref{table2} and Figures
\ref{optimal_number}, \ref{sharpe} for FF48.

From the results of our two exercises we see that:
\begin{itemize}
\item Our sparse portfolios (with a relatively small number of
assets and moderate $\tau$) outperform the na\"{i}ve $1/N$ strategy
significantly and consistently over the entire evaluation period.
This gain is achieved for a wide range of portfolio sizes, as
indicated in Figures \ref{sharpe} and \ref{sharpeFF100}. It is to be
noted that the best performing sparse portfolio we constructed is
{\em not} always the no-short-positions portfolio.

\item When we target a large number of assets in our portfolio,
the performance deteriorates. We interpret this as a result of
so-called ``overfitting". Larger target numbers of assets correspond
to smaller values of $\tau$. The $\ell_1$ penalty is then having
only a negligible effect and the minimization focuses essentially on
the variance term. Hence, the solution becomes unstable and is
overly sensitive to the estimation errors that plague the original
(unpenalized) Markowitz optimization problem (\ref{regression}).
\end{itemize}

\begin{table}
\caption{Performance of the sparse portfolio with no short-selling,
for FF100.}

\centering
\begin{tabular}{|c||c|c|c||c|c|c|}
\multicolumn{7}{c}{}\\
\hline Evaluation&\multicolumn{3}{c||}{$w_i \geqslant 0$}&
\multicolumn{3}{c|}{Equal} \\
period&\multicolumn{3}{c||}{for all $i$}&\multicolumn{3}{c|}{weight} \\
&$m$&$\sigma$&$S$&$m$&$\sigma$&$S$\\
\hline \hline
06/76-06/06 &   16 & 53 & 30 &   17 & 59 & 28\\
\hline
07/76-06/81  &  12 & 59 & 21 &   23 & 61 & 38\\
07/81-06/86  &  24 & 49 & 49 &   20 & 53 & 38\\
07/86-06/91  &  10 & 65 & 15 &    9 & 71 & 13\\
07/91-06/96  &  19 & 31 & 61 &   18 & 34 & 53\\
07/96-06/01  &  18 & 52 & 35 &   16 & 63 & 26\\
07/01-06/06  &  11 & 55 & 21 &   12 & 64 & 19\\
\hline
\end{tabular}
\parbox{9cm}{\footnotesize Monthly mean return $m$, standard deviation of monthly return
$\sigma$, and corresponding Sharpe ratio $S$ (expressed in \%) for
the optimal positive-weights-only portfolio, as well as for the
$1/N$-strategy portfolio. Both portfolio strategies are tested for
their performance over twelve consecutive months immediately
following their construction; their returns are pooled over several
years to assess their performance, as described in the text.
Statistics reflect the performance an investor would have achieved,
on average, by constructing the portfolio on July 1 one year, and
keeping it for the next twelve months, until June 30 of the next
year; the average is taken over 5 years for each of the break-out
periods, over 30 years for the full period.}
\label{table3}
\end{table}

\begin{table}
\caption{Empirical results for FF100.}

\resizebox{\textwidth}{!}{
\begin{tabular}{|c|c|c|c|c|c|c|c|c|c|c|c|c|c|c|c|c|c|c|c|c|c|}
\hline
Evaluation&\multicolumn{3}{c|}{bin=11-20}&\multicolumn{3}{c|}{bin=21-30}&\multicolumn{3}{c|
}{bin=31-40}&\multicolumn{3}{c|}{bin=41-50}&\multicolumn{3}{c|}{bin=51-60}&\multicolumn{3}{c|}{$w_i
\geqslant 0$} \\
period&$m$&$\sigma$&$S$&m&$\sigma$&$S$&$m$&$\sigma$&$S$&$m$&$\sigma$&$S$&$m$&$\sigma$&$S$&$m$&$\sigma$&$S$\\
\hline
\!\!06/76-06/06\!\!&\!\!16\!\!&\!\!50\!\!&\!\!33\!\!&\!\!19\!\!&\!\!48\!\!&\!\!39\!\!&\!\!19\!\!&\!\!49\!\!&\!\!40\!\!&\!\!20\!\!&\!\!52\!\!&\!\!39\!\!&\!\!21\!\!&\!\!60\!\!&\!\!34\!\!&\!\!16\!\!&\!\!53\!\!&\!\!30\!\!\\
\hline
\!\!07/76-06/81\!\!&\!\!11\!\!&\!\!52\!\!&\!\!22\!\!&\!\!12\!\!&\!\!51\!\!&\!\!24\!\!&\!\!12\!\!&\!\!55\!\!&\!\!22\!\!&\!\!11\!\!&\!\!56\!\!&\!\!20\!\!&\!\!7\!\!&\!\!66\!\!&\!\!10\!\!&\!\!12\!\!&\!\!59\!\!&\!\!21\!\!\\
\!\!07/81-06/86\!\!&\!\!26\!\!&\!\!41\!\!&\!\!64\!\!&\!\!31\!\!&\!\!38\!\!&\!\!81\!\!&\!\!31\!\!&\!\!40\!\!&\!\!77\!\!&\!\!31\!\!&\!\!43\!\!&\!\!72\!\!&\!\!33\!\!&\!\!49\!\!&\!\!67\!\!&\!\!24\!\!&\!\!49\!\!&\!\!49\!\!\\
\!\!07/86-06/91\!\!&\!\!9\!\!&\!\!63\!\!&\!\!14\!\!&\!\!9\!\!&\!\!61\!\!&\!\!16\!\!&\!\!11\!\!&\!\!62\!\!&\!\!18\!\!&\!\!12\!\!&\!\!64\!\!&\!\!19\!\!&\!\!12\!\!&\!\!71\!\!&\!\!17\!\!&\!\!10\!\!&\!\!65\!\!&\!\!15\!\!\\
\!\!07/91-06/96\!\!&\!\!20\!\!&\!\!29\!\!&\!\!70\!\!&\!\!22\!\!&\!\!25\!\!&\!\!86\!\!&\!\!20\!\!&\!\!28\!\!&\!\!73\!\!&\!\!22\!\!&\!\!31\!\!&\!\!70\!\!&\!\!25\!\!&\!\!36\!\!&\!\!67\!\!&\!\!19\!\!&\!\!31\!\!&\!\!61\!\!\\
\!\!07/96-06/01\!\!&\!\!18\!\!&\!\!53\!\!&\!\!35\!\!&\!\!23\!\!&\!\!52\!\!&\!\!44\!\!&\!\!29\!\!&\!\!47\!\!&\!\!61\!\!&\!\!31\!\!&\!\!50\!\!&\!\!62\!\!&\!\!34\!\!&\!\!63\!\!&\!\!54\!\!&\!\!18\!\!&\!\!52\!\!&\!\!35\!\!\\
\!\!07/01-06/06\!\!&\!\!11\!\!&\!\!53\!\!&\!\!22\!\!&\!\!15\!\!&\!\!51\!\!&\!\!29\!\!&\!\!13\!\!&\!\!51\!\!&\!\!26\!\!&\!\!13\!\!&\!\!56\!\!&\!\!23\!\!&\!\!14\!\!&\!\!64\!\!&\!\!22\!\!&\!\!11\!\!&\!\!55\!\!&\!\!21\!\!\\
\hline
\end{tabular}}

\parbox{\textwidth}{Monthly mean return, $m$, standard deviation of monthly return,
$\sigma$, and corresponding monthly Sharpe ratio $S$ (expressed in
\%) for the optimal portfolios with 11-20, 21-30, 31-40, 41-50,
51-60 assets, as well as (again) the optimal portfolio without short
positions. Portfolios are constructed annually as described in the
text. Statistics reflect the performance an investor would have
achieved on average, by constructing the portfolio on July 1 one
year, and keeping it for the next twelve months, until June 30 of
the next year; the average is taken over several years: 5 for each
break-out period, 30 for the full period.}

\label{table4}\label{TableBins100}
\end{table}

\section{Possible generalizations}

In this section, we briefly describe some extensions of our
approach. It should be pointed out that the relevance and usefulness
of the $\ell_1$ penalty is not limited to a stable implementation of
the usual Markowitz portfolio selection scheme described in Section
2. Indeed, there are several other portfolio construction problems
that can be cast in similar terms or otherwise solved through the
minimization of a similar objective function. We now list a few
examples:

\subsection{Partial index tracking}

In many situations, investors want to create a portfolio that
efficiently tracks an index.  In some cases, this will be an
existing financial index whose level is tied to a large number of
tradable securities but which is not yet tradable en masse as an
index future or other single instrument.  In such a situation,
investors need to find a collection of securities whose profit and
loss profile accurately tracks the index level.  Such a collection
need not be a full replication of the index in question;  indeed
it is frequently inconvenient or impractical to maintain a full
replication.

In other situations, investors will want to monetize some more
abstract financial time series: an economic time series, an
investor sentiment time series, etc.  In that case, investors will
need to find a collection of securities which is likely to remain
correlated to the target time series.

Either way, the investor will have at his disposal a time series
of index returns, which we will write as a $T \times 1$ column
vector, $\y$.  Also, the investor will have at his disposal the
time series of returns for every available security, which we will
write as a $T \times N$ matrix $\pmb{R}$, as before.

In that case, an investor seeking to minimize expected tracking
error would want to find
\begin{equation*}
\widehat{\w} = \arg \min_{\w}  \|\y-\pmb{R}\w\|_2^2.
\end{equation*}
However, this problem is simply a linear regression of the target
returns on the returns of the available assets.  As the available
assets may be collinear, the problem is subject to the same
instabilities that we discussed above.  As such, we can augment our
objective function with an $\ell_1$ penalty and seek instead
\begin{equation*}
\w^{[\tau]} = \arg \min_{\w} \left[ \|\y-\pmb{R}\w\|_2^2 + \tau
\|\w\|_1\right]~,
\end{equation*}
subject to the appropriate constraints. This simple modification
stabilizes the problem and enforces sparsity, so that the index
can be stably replicated with few assets.

Moreover, one can enhance this objective function in light of the
interpretation of the $\ell_1$ term as a model of transaction costs.
Let $s_i$ is the transaction cost (bid-ask spread) for the $i$th
security.  In that case, we can seek
\begin{equation*}
\w^{[\tau]} = \arg \min_{\w} \left[ \|\y-\pmb{R}\w\|_2^2 + \tau
\sum_i s_i |w_i|\right]~.
\end{equation*}
By making this modification, the optimization process will ``prefer"
to invest in more liquid securities (low $s_i$) while it will
``avoid" investments in less liquid securities (high $s_i$). A
slightly modified version of the algorithm described above can cope
with such weighted $\ell_1$ penalty and generate a list of
portfolios for a wide range of values for $\tau$.  For each
portfolio, the investor could then compare the expected tracking
error per period ($\frac{1}{\,^{\,}T\,} \|\y-\pmb{R}\w\|_2^2$) with
the expected cost of creating and liquidating the tracking portfolio
($\sum_i s_i |w_i|$).  The investor could then select a portfolio
that suits both his risk tolerance and cost constraints.

\subsection{Portfolio hedging}

Consider the task of hedging a given portfolio using some subset
of a universe of available assets.  As a concrete example, imagine
trying to efficiently hedge out the market risk in a portfolio of
options on a single underlying asset, potentially including many
strikes and maturities.  An investor would be able to trade the
underlying asset as well as any options desired.  In this context,
it would be possible to completely eliminate market risk by
negating the initial position.  However, this may not be feasible
given liquidity (transaction cost) constraints.

Instead, an investor may simply want to reduce his risk in a
cost-efficient way.  One could proceed as follows: Generate a list
of scenarios.  For each scenario, determine the change in the
value of the existing portfolio.  Also, determine the change in
value for a unit of each available security.  Store the former in
a $M \times 1$ column vector $\y$ and store the latter in a
$M \times N$ matrix, $\pmb{X}$.  Also, determine a
probability, $p_i$ for $i = 1, \ldots, M$ of each scenario, and
store {\em the square root of these values} in a diagonal $M \times M$ matrix, $\pmb{P}$.
These probabilities can be derived from the market or assumed
subjectively according to an investor's preference.  As before,
denoting by $s_i$ the transaction costs for each tradable
security, we can seek
\begin{equation*}
\w^{[\tau]} = \arg \min_{\w} \left[
\|\pmb{P}\left(\y+\pmb{X}\w\right)\|_2^2 + \tau \sum_i s_i
|w_i|\right]~.
\end{equation*}
As before, the investor could then apply one of the algorithms
above to generate a list of optimal portfolios for a wide range of
values of $\tau$.  Then, just as in the index tracking case, the
investor could observe the attainable combinations of expected
mark to market variance ($\|\pmb{P}\left(\y+\pmb{X}\w\right)\|_2^2$) and
transaction cost ($\sum_i s_i |w_i|$).  One appealing feature of
this method is that it does not explicitly determine the number of
assets to be included in the hedge portfolio.  The optimization
naturally trades off portfolio volatility for transaction cost,
rather than imposing an artificial cap on either.

\subsection{Portfolio Adjustment}

Thus far, we have assumed that investors start with no assets, and
must construct a portfolio to perform a particular task.  However,
this is rarely the case in the real world.  Instead, investors
frequently hold a large number of securities and must modify their
existing holdings to achieve a particular goal.  In this context,
the investor already holds a portfolio $\w$ and must make an
adjustment $\Delta_\w$.  In that case, the final portfolio will be
$\w + \Delta_\w$, but the transaction costs will be relevant only
for the adjustment $\Delta_\w$. The corresponding optimization
problems is given by
\begin{eqnarray}
\Delta_\w^{[\tau]} & = &
\arg \min_{\Delta_\w}\left[  \|\rho\1_T-\pmb{R}(\w+\Delta_\w)\|^2_2
+ \tau\| \Delta_\w\|_1\right]\nonumber\\
& \mbox{s. t.} & \Delta_\w\trp\widehat{\mmu}= 0 \nonumber\\
& & \Delta_\w\trp \1_N= 0.\nonumber
\end{eqnarray}
It is easy to modify our methodology to handle this situation.

\section{Conclusion}

We have devised a method that constructs stable and sparse
portfolios by introducing an $\ell_1$ penalty in the Markowitz
portfolio optimization. We obtain as special cases the
no-short-positions portfolios which also comprise few active assets.
To our knowledge, such a sparsity property of the non-negative
portfolios has not been previously noticed in the literature. The
portfolios we propose can be seen as natural extensions of the
no-short-positions portfolios and maintain or improve their
performances while preserving as much as possible their sparse
nature.

We have also described an efficient algorithm for computing the
optimal, sparse portfolios, and we have implemented it using as
assets two sets of portfolios constructed by Fama and French: 48
industry portfolios and 100 portfolios formed on size and
book-to-market. We found empirical evidence that the optimal
sparse portfolios outperform the evenly-weighted portfolios by
achieving a smaller variance; moreover they do so with only a
small number of active positions, and the effect is observed over
a range of values for this number. This shows that adding an
$\ell_1$ penalty to objective functions is a powerful tool for
various portfolio construction tasks. This penalty forces our
optimization scheme to select, on the basis of the training data,
few assets forming a stable and robust portfolio, rather than
being ``distracted'' by the instabilities due to collinearities
and responsible for meaningless artifacts in the presence of
estimation errors.

Many variants and improvements are possible on the simple procedure
described and illustrated above. This goes beyond the scope of the
present paper which was to propose a new methodology and to
demonstrate its validity. In future work, we plan to extend our
empirical exercises to other and larger asset collections (such as
S$\&$P500), to explore other performance criteria than the usual
Sharpe ratio, and to derive automatic procedures for choosing the
number of assets to be included in our portfolio. We believe that
the good regularization properties of our method should ensure its
robustness against these various changes.

\section*{Acknowledgments}
We thank Tony Berrada, Laura Coroneo, Simone Manganelli, Sergio
Pastorello, Lucrezia Reichlin and Olivier Scaillet for helpful
suggestions and discussions. Part of this research has been
supported by the ``Action de Recherche Concert\'ee'' Nb 02/07-281
(CDM and DG), the Francqui Foundation (IL), the VUB-GOA 62 grant
(ID, CDM and IL), the National Bank of Belgium BNB (DG and CDM), and
by the NSF grant DMS-0354464 (ID). Joshua Brodie thanks Bobray
Bordelon, Jian Bai, Josko Plazonic and John Vincent for their
valuable technical assistance during his senior thesis work, out of
which grew his
and ID's participation in this project.\\
The opinions in this paper are those of the authors and do not
necessarily reflect the views of the European Central Bank.

\bibliographystyle{econometrica}
\bibliography{biblio}

\appendix

\section*{Appendix: Constrained minimization algorithm}

Before discussing our solution method for the linearly constrained
$\ell_1$-penalized least-squares problem, we briefly recall the
homotopy/LARS method which manages to recover the unconstrained
minimizer of the $\ell_1$-penalized least-squares objective function
\begin{displaymath}
\bar \w(\tau)=\arg\min_\w \left[\|\R \w-\y\|_2^2+\tau\|\w\|_1\right]
\end{displaymath}
for a whole range of values of the (positive) penalty parameter
$\tau$.

The variational equations describing the minimizer $\bar \w(\tau)$
are:
\begin{eqnarray}
\displaystyle (\R \trp(\y-\R \w))_i &=& \frac{\tau}{2}\, \mathrm{sgn}\, w_i \qquad\qquad w_i\neq 0\label{kkt1}\\
\displaystyle |(\R \trp(\y-\R \w))_i|&\leq&\frac{\tau}{2}
\qquad\qquad\qquad\,\,\,\, w_i= 0.\label{kkt2}
\end{eqnarray}
The minimizer $\bar \w(\tau)$ is a continuous piecewise linear
function of $\tau$. We shall denote the breakpoints by
$\tau_0>\tau_1>\ldots$ and the corresponding minimizers by $\bar
\w(\tau_0), \bar \w(\tau_1), \ldots$ The breakpoints occur where a
new component enters or leaves the support of $\bar \w(\tau)$. We
will use $\bb$ to denote the residual $\bb=\R \trp (\y-\R \w)$.

The homotopy/LARS method for solving these equations starts by
considering the point $\w=0$, which satisfies  the equations
(\ref{kkt1},\ref{kkt2}) for all $\tau\geq\tau_0\equiv 2\max_i|(\R
\trp \y)_i|$. Hence $\bar \w(\tau\geq \tau_0)=0$.

Given a breakpoint $\bar \w(\tau_n)$, it is possible to construct
the next breakpoint $\bar \w(\tau_{n+1})$ by solving a small linear
system. Let $J=\{i \mathrm{\ for\ which\ }
\left|\bb_i\right|=\tau_{n}/2 \}$ (i.e. the set of maximal
residual), $\R _J$ the submatrix consisting of the columns $J$ of
$\R $. We define the walking direction $\uu$ by
\begin{displaymath}
\R _J\trp \R _J\, \uu_J=\mathrm{sgn}\left(\bb_J\right)
\end{displaymath}
and $u_i=0$ for $i\not\in J$ ($\mathrm{sgn}\left(\bb_J\right)$
denotes the vector $(\mathrm{sgn}(b_j)_{j\in J})$). In this way, a
step $\w\rightarrow \w+\gamma \uu$ results in a change in the
residual $\bb\rightarrow \bb-\gamma \vv$, where
$v_j=\mathrm{sgn}(b_j)$ for $j\in J$. In other words, the maximal
components of the residual decrease at the same rate. The step size
$\gamma>0$ is now determined to be the smallest number for which the
absolute value of a component $|b_i-\gamma v_i|$ (with $i\not\in J$)
of the new residual becomes equal to $|b_j-\gamma v_j|$ for $j\in J$
(i.e. a new component joins the maximal residual set), or for which
a nonzero component of $\w$ is turned into zero.

The new penalty parameter is then $\tau_{n+1}=\tau_{n}-2\gamma$
(which is smaller than $\tau_{n}$), and the corresponding minimizer
is $\bar \w(\tau_{n+1})=\bar \w(\tau_{n})+\gamma \uu$. By
construction it is guaranteed to satisfy the variational equations
(\ref{kkt1},\ref{kkt2}).

The two main advantages of this method are thus that it is exact (in
particular zero components are really zero) and that it yields the
breakpoints (and hence the minimizers) for a whole range of values
of the penalization parameters $\tau\geq\tau_\mathrm{stop}\geq0$. At
each step, only a relatively small linear system has to be solved.
If this procedure is carried through until the end, one finds
$\lim_{\tau\rightarrow 0}\arg\min\|\R \w-\y\|_2^2+\tau\|\w\|_1
=\displaystyle \arg\!\!\!\!\!\!\!\!\!\!\!\!\!\min_{\w\
\mathrm{s.t.}\ \R \trp \R \w=\R \trp \y}\|\w\|_1$.

For the constrained case, i.e. the minimization problem
\begin{equation}
\widetilde \w(\tau)=\arg\!\!\!\!\!\!\!\!\!\min_{\w \ \mathrm{s.t.}\
\A\w=\aa} \left[\|\R
\w-\y\|_2^2+\tau\|\w\|_1\right]\label{constrl1problem}
\end{equation}
subject to the linear constraint $\A\w=\aa$, we can devise a similar
procedure. We assume, of course, that the constraint $\A\w=\aa$ has a
solution.

An approximation of the minimizer $\widetilde \w(\tau)$ can be
obtained by applying the unconstrained procedure described above to
the objective function
\begin{equation}
\widetilde{\widetilde{\w}}(\tau_\epsilon)=\arg\min_\w\left[\|\A\w-\aa\|_2^2+\epsilon\|\R
\w-\y\|_2^2+\tau_\epsilon\|\w\|_1\right].\label{functionaleps}
\end{equation}
For sufficiently small $\epsilon$, this will give a good
approximation of the constrained minimizer $\widetilde \w(\tau)$
corresponding to the penalty $\tau=\tau_\epsilon/\epsilon$ (after
first going through a number of breakpoints for which $\A\w\neq
\aa$, not even approximately). However, this is clearly an
approximate method (often very good) whereas the unconstrained
procedure did not involve any approximation.

We solve this issue, and provide an exact method, by solving the
minimization problem (\ref{functionaleps}) up to the first order in
$\epsilon$. In this approach $\epsilon$ is a small \emph{formal}
positive parameter. Now the minimizer $\widetilde{\widetilde{\w}}
(\tau_\epsilon)$ and $\tau_\epsilon$ both depend on $\epsilon$. We
can write $\w=\w^{(0)}+\epsilon \w^{(1)}+\mathcal{O}(\epsilon^2)$
and $\tau_\epsilon=\tau^{(0)}+\tau^{(1)}
\epsilon+\mathcal{O}(\epsilon^2)$. We again follow the procedure for
the unconstrained method, but take care to use arithmetic (addition,
multiplication, comparison, \ldots) up to first order in $\epsilon$.

As before, one starts from $\w=0$, corresponding to a large initial
value of $\tau_\epsilon$, and follows the path of descending
$\tau_\epsilon$. The strategy consists of satisfying the variational
equations
\begin{eqnarray}
\displaystyle \left(\A\trp (\aa-\A\w)+\epsilon \R \trp (\y-\R \w)\right)_i &=
& \frac{\tau_\epsilon}{2}\, \mathrm{sgn}\, w_i \qquad\qquad w_i\neq 0\label{kkttris1}\\
\displaystyle \left|\left(\A\trp (\aa-\A\w)+\epsilon \R \trp (\y-\R
\w)\right)_i\right|&\leq&\frac{\tau_\epsilon}{2} \qquad\qquad\qquad
\,\,\,\,w_i= 0\label{kkttris2}
\end{eqnarray}
at each breakpoint by carefully determining a walking direction
$\uu=\uu^{(0)}+\uu^{(1)} \epsilon+\mathcal{O}(\epsilon^2) $ and a
step length $\gamma=\gamma^{(0)}+\gamma^{(1)}
\epsilon+\mathcal{O}(\epsilon^2)$. Using $\w=\w^{(0)}+\epsilon
\w^{(1)}+\mathcal{O}(\epsilon^2)$, we can rewrite equations
(\ref{kkttris1}) as
\begin{eqnarray}
\displaystyle \left(\A\trp (\aa-\A\w^{(0)})\right)_i &=& \frac{\tau^{(0)}}{2}\, \mathrm{sgn}\, w_i \label{eq1}\\
\displaystyle \left(-\A\trp \A\w^{(1)}+\R \trp (\y-\R
\w^{(0)})\right)_i &=& \frac{\tau^{(1)}}{2}\, \mathrm{sgn}\, w_i\ .
\label{eq2}
\end{eqnarray}
From a known breakpoint $\w$ we can proceed to the following
breakpoint by a step direction $\uu$ and step size $\gamma$ (both
depending on $\epsilon$). We again set
\begin{displaymath}
J=\arg\max_i\left|\left(\A\trp (\aa-\A\w^{(0)})+\epsilon (-\A\trp
\A\w^{(1)}+\R \trp (\y-\R \w^{(0)}))\right)_i\right|.
\end{displaymath}
As long as $\tau^{(0)}\neq 0$, the components $J$ of $\uu$ are
determined by
\begin{equation}
\left(
\begin{array}{cc}
\R _J\trp \R _J & \A_J\trp \A_J\\
\A_J\trp \A_J & 0
\end{array}
\right) \left(
\begin{array}{c}
\uu_{J}^{(0)}\\
\uu_{J}^{(1)}
\end{array}
\right) = \left(
\begin{array}{c}
0\\
\mathrm{sgn}(\bb_J)
\end{array}
\right)\label{linsyseps}
\end{equation}
and the other components of $\uu$ remain zero. The step size
$\gamma$ is again determined as before, i.e. when a new component
enters the maximal residual set, or when a component leaves the
active set. The penalty parameter $\tau_\epsilon$ decreases as
before: $\tau_\epsilon\rightarrow \tau_\epsilon-2\gamma$.

At some point in this procedure, $\tau_\epsilon$ will become zero in
zeroth order:
$\tau_\epsilon=0+\tau^{(1)}\epsilon+\mathcal{O}(\epsilon^2)$. The
corresponding minimizer (more precisely the zeroth-order part of
this breakpoint) will satisfy the constraint $\A\w=\aa$ and we will
have found the first constrained minimizer $\widetilde \w$ of
(\ref{constrl1problem}), corresponding to $\tau_0=\tau^{(1)}$ (i.e.
the first-order part of the parameter $\tau_\epsilon$ of the
$\epsilon$-dependent problem at this breakpoint). In the
unconstrained case, no such calculations were necessary as the
starting point was always equal to $0$. Similarly to the
unconstrained case, we have that $\widetilde \w(\tau >
\tau_0)=\widetilde \w(\tau_0)$.

In principle, one could continue the $\epsilon$-dependent algorithm,
but now that the first breakpoint of $\widetilde \w(\tau)$ is
determined, it is more advantageous to continue the descent of
$\tau$ by introducing Lagrange multipliers $\llambda$ for the
problem (\ref{constrl1problem}):
\begin{displaymath}
\widetilde \w(\tau)=\arg\!\!\!\!\!\!\!\!\!\!\!\min_{\llambda,\,\,
\w\ \mathrm{s.t.}\ \A\w=\aa} \left[\|\R
\w-\y\|_2^2+\tau\|\w\|_1+2\llambda\trp (\A\w-\aa)\right].
\end{displaymath}
This minimization problem (analogous to (\ref{constrl1problem}))
amounts to solving the equations:
\begin{eqnarray}
\displaystyle (\R \trp (\y-\R \w)+\A\trp \llambda)_i &=& \frac{\tau}{2}\, \mathrm{sgn}\, w_i \qquad\qquad w_i\neq 0\label{kktbis1}\\
\displaystyle |(\R \trp (\y-\R \w)+\A\trp
\llambda)_i|&\leq&\frac{\tau}{2} \qquad\qquad\qquad\,\,\,\, w_i=
0\label{kktbis2}\\
\A\w&=&\aa.\label{kktbis3}
\end{eqnarray}
Equation (\ref{kktbis1}) is the equivalent of equation (\ref{eq2})
whereas equation (\ref{kktbis3}) replaces equation (\ref{eq1}). We
now already have $\tau_0$, $\widetilde \w(\tau\geq \tau^{(0)})$ and
the initial Lagrange multipliers
$\llambda=-\A\widetilde{\widetilde{\w}}^{(1)}$ (from the first-order
part of the last step of the $\epsilon$-dependent problem).

To proceed from one breakpoint to the next ($\w\rightarrow \w+\gamma
\uu$, and $\llambda\rightarrow \llambda+\gamma \sss$ as the
multipliers also change), we again need to solve a linear system:
\begin{equation}
\left(
\begin{array}{cc}
\R _J\trp \R _J & \A_J\trp\\
\A_J &0
\end{array}
\right) \left(
\begin{array}{c}
\uu_J\\
\sss
\end{array}
\right) = \left(
\begin{array}{c}
\mathrm{sgn}(\tilde \bb_J)\\
0
\end{array}
\right)\label{linsysconstr}
\end{equation}
with $\tilde \bb=\R \trp (\y-\R \w)+\A\trp \llambda$. This will
guarantee that $\w\rightarrow \w+\gamma \uu$ and
$\llambda\rightarrow \llambda+\gamma \sss$ still satisfy the
constraint (\ref{kktbis3}) and the variational equations
(\ref{kktbis1},\ref{kktbis2}). The step size $\gamma$ is determined
by the same rule as before: stop when a new component enters the set
$J=\arg\max_i |(\R \trp (\y-\R \w)+\A\trp \llambda)_i|$ or when a
nonzero component of $\w$ is set to zero. Notice the differences and
similarities between the linear systems (\ref{linsysconstr}) and
(\ref{linsyseps}).

At each breakpoint, this algorithm provides the penalty $\tau_n$,
the corresponding minimizer $\widetilde \w(\tau_n)$ and the Lagrange
multipliers $\llambda_n$. Unlike for the unconstrained case, it is
now possible that $\widetilde \w(\tau)$ remains constant between two
breakpoints (i.e. only the Lagrange multipliers $\llambda$ change).

One simplifying assumption (not solved in the homotopy/LARS
algorithm) was made in the above description of the algorithm: if
the set of maximal residual and the support set differ by more than
one component, one should carefully select the correct new
components to enter the support. This can be done by using the
variational equations, and our implementation handles this case.

One could argue that the starting point (i.e. the first breakpoint)
for the constrained minimization problem is simply given by
$\widetilde \w(\tau_0)=\displaystyle\arg\!\!\!\!\!\!\!\!\!\min_{\w\
\mathrm{s.t.}\ \A\w=\aa} \|\w\|_1$, which could be calculated by
letting the unconstrained solution procedure run its course:
$\widetilde \w(\tau_0)=\lim_{\sigma\rightarrow 0}\arg\min_\w\left[
\|\A\w-\aa\|_2^2+\sigma\|\w\|_1\right]$. Generically (i.e. excluding
special cases), this is correct. However, the problem is that
sometimes the minimizer $\arg\!\min_{\w\ \mathrm{s.t.}\ \A\w=\aa}
\|\w\|_1$ is not unique. In that case, the starting point for the
constrained minimizer is not solely determined by $\A$ and $\aa$ but
also by $\R $ and $\y$. In this case, the $\epsilon$-dependent
algorithm still chooses the correct starting point from the set
$\displaystyle \arg\!\!\!\!\!\!\!\!\min_{\w\ \mathrm{s.t.}\
\A\w=\aa}\|\w\|_1$. This is important to mention because the special
constraint $\sum w_i=1$ used in this paper, gives rise to such
cases.

Our algorithm is well-suited for the portfolio problems discussed in
this paper. The size of the matrix, the number of constraints (just
two) and, more importantly, the number of nonzero weights in the
portfolios are such that a minimization run (i.e. finding the
minimizer for a whole range of penalty parameters) can be done in a
fraction of a second on a standard desktop.

We calculated the portfolio examples in this paper using both the
formal $\epsilon$ approach (in Mathematica) and the approximate
small $\epsilon$ approach (in Matlab). The outcomes were always
consistent.

\end{document}